\pdfoutput=1
\documentclass[preprint,11pt,number,3p,times]{elsarticle}
\usepackage[utf8,latin1]{inputenc} 

\bibstyle{elsarticle-num}
\usepackage[centertags]{amsmath}
\usepackage{amssymb}
\usepackage{epsfig}
\usepackage{varioref}
\usepackage{xspace}
\usepackage{bbm}
\usepackage{lscape}
\usepackage{paralist}
\usepackage{listings}
\usepackage{caption}
\def\CC{{C\nolinebreak[4]\hspace{-.05em}\raisebox{.4ex}{\tiny\bf ++}}\xspace}
\DeclareCaptionFont{white}{\color{white}}
\DeclareCaptionFormat{listing}{%
  \parbox{\textwidth}{\colorbox{gray}{\parbox{\textwidth}{#1#2#3}}\vskip-4pt}}
\usepackage[T1]{fontenc}
\usepackage[scaled]{beramono}

\usepackage{textcomp}

\usepackage[usenames,dvipsnames,table]{xcolor}
\usepackage{fancyvrb}
\usepackage{microtype}

\def\CC{{C\nolinebreak[4]\hspace{-.05em}\raisebox{.4ex}{\tiny\bf ++}}\xspace}
\newcommand{\code}[1]{{{\tt #1}}}
\newcommand{\codeq}[1]{\textquotedbl{\tt #1}\textquotedbl\xspace}
\newcommand{\pyrate}{PyR@TE\xspace}
\newcommand{\pylie}{PyLie\xspace}
\newcommand{\pyratetwo}{PyR@TE\xspace2\xspace}
\newcommand{\python}{\code{Python}\xspace}
\newcommand{\sarah}{\code{SARAH}\xspace}
\newcommand{\mathematica}{\code{Mathematica}\xspace}

\newcommand{\SU}[1]{\ensuremath{\mathrm{SU}(#1)}}
\newcommand{\U}[1]{\ensuremath{\mathrm{U}(#1)}}

\newcommand{\rep}[1]{\ensuremath{\boldsymbol{#1}}}
\newcommand{\crep}[1]{\ensuremath{\overline{\boldsymbol{#1}}}}

\usepackage[colorlinks=true
,urlcolor=blue
,anchorcolor=blue
,citecolor=blue
,filecolor=blue
,linkcolor=black
,menucolor=blue
,linktocpage=true
]{hyperref}

\labelformat{section}{Section~#1}
\labelformat{subsection}{Section~#1}
\labelformat{equation}{Eq.~(#1)}
\labelformat{figure}{Fig.~#1}
\labelformat{subfigure}{Fig.~\thefigure#1}
\labelformat{table}{Tab.~#1}

\journal{Computer Physics Communications}

\AtBeginDocument{\labelformat{lstlisting}{Listing~#1}}

\begin{document}

\begin{frontmatter}

%\titlehead{
\begin{flushright}
\normalsize{LPSC-16-170}\\
\normalsize{SMU-16-11}
\end{flushright}
%}

\title{{\huge\bfseries PyR@TE 2}\\A Python tool for computing RGEs at two-loop.}

\author[smu]{F.~Lyonnet\footnote{Corresponding author: florian.lyonnet@lpsc.in2p3.fr}}
\author[lpsc]{I.~Schienbein}

\address[smu]{Southern Methodist University, Dallas, TX 75275, USA}
\address[lpsc]{Laboratoire de Physique Subatomique et de Cosmologie, UJF Grenoble 1, CNRS/IN2P3, INPG,\newline 53 Avenue des Martyrs, F-38026 Grenoble, France}

\begin{abstract}
	Renormalization group equations are an essential tool for the description of theories accross different energy scales.
	Even though their expressions at two-loop for an arbitrary gauge field theory have been known for more than thirty years, deriving the full set of equations for a given model by hand is very challenging and prone to errors. To tackle this issue, we have introduced in \cite{Lyonnet:2013dna} a \python tool called \pyrate; {\it Python Renormalization group equations @ Two-loop for Everyone}. With \pyrate, it is easy to implement a given Lagrangian and derive the complete set of two-loop RGEs for all the parameters of the theory. In this paper, we present the new version of this code, \pyratetwo, which brings many new features and in particular it incorporates kinetic mixing when several \U{1} gauge groups are involved. In addition, the group theory part has been greatly improved as we introduced a new \python module dubbed PyLie that deals with all the group theoretical aspects required for the calculation of the RGEs as well as providing very useful model building capabilities. This allows the use of any irreducible representation of the $\mathrm{SU}(n)$, $\mathrm{SO}(2n)$ and $\mathrm{SO(2n+1)}$ groups. 
Furthermore, it is now possible to implement terms in the Lagrangian involving fields which can be contracted into gauge singlets in more than one way. As a byproduct, results for a popular model (SM+complex triplet) for which, to our knowledge, the complete set of two-loop RGEs has not been calculated before are presented in this paper. 
Finally, the two-loop RGEs for the anomalous dimension of the scalar and fermion fields have been implemented as well. It is now possible to export the coupled system of beta functions into a numerical \CC function, leading to a consequent speed up in solving them.
	
\end{abstract}

\end{frontmatter}
\section{Introduction}
% ADD a general sentence from Fred
The renormalization group equations (RGEs) for general non-supersymmetic gauge theories have been 
known at two-loop accuracy for more than 30 years~\cite{Machacek:1983tz,Machacek:1983fi,Machacek:1984zw,Jack:1982hf,Jack:1982sr,Jack:1984vj}. 
Due to their importance, these results have been thoroughly scrutinized in the literature and a
complete re-evaluation has been performed in \cite{Luo:2002ti}.
The general RGEs have been implemented in the \python program \pyrate~\cite{Lyonnet:2013dna}\footnote{The code and various tutorials are publicly available at \url{http://pyrate.hepforge.org}.}
which automatically generates the full two-loop RGEs for all the (dimensionless and dimensionful) parameters 
of a general gauge theory 
once the gauge group and particle content have been specified by the user via simple text files.
All known typos in the original series of papers by Machacek and Vaughn have been taken into account 
(see, e.g., \cite{Luo:2002ti} and the appendix of Ref.\ \cite{Wingerter:2011dk}) and the code has been extensively validated against the literature. In addition, independently of \pyrate, \mathematica routines were developed and cross-checked against \pyrate; these routines are now part of \sarah 4~\cite{Staub:2013tta}. 

The code was designed to be easily usable and to provide the user with maximum flexibility for a variety of calculations. 
For example, one can choose to calculate the $\beta$-function of only one of the terms defined in the potential or even neglect a specific contribution to a particular RGE. 
These options are particularly useful for quick validations when developing a model, or to avoid calculating terms 
that are time consuming and that will be neglected in the end. 
The full list of options can be found in \ref{tab:options} and will be further described below.
Once the RGEs have been calculated by \pyrate the results can be exported to \LaTeX\ and \mathematica\ or 
stored in a \python data structure for further processing. 
In addition, it is possible to create a numerical \python function that can be used to solve the RGEs, e.g. using the provided \python tool box\footnote{Note that a \mathematica routine to solve the exported RGEs is also provided.}.

Since the first release of \pyrate~\cite{Lyonnet:2013dna}  three years ago,  
the code has been improved and extended by several new features described in this article
which warrants the release of a new version 2 of \pyrate\ :
\begin{itemize}
\item 
%A main new feature of \pyrate 2 is the capability to handle semi-simple gauge groups
\pyrate 2 is now able to handle semi-simple gauge groups
with more than one Abelian gauge factor including the mixing between the associated gauge kinetic
fields. The corresponding two-loop $\beta$-functions for the gauge couplings were studied in
\cite{delAguila:1988jz,delAguila:1987st} 
and the general two-loop $\beta$-functions for all other dimensionless parameters were 
derived in \cite{Luo:2002iq}. 
Our implementation employs the method proposed in 
Ref.\  \cite{Fonseca:2013jra} where also the two-loop $\beta$-functions 
for the dimensionful parameters can be found.
We have taken into account all the corrections at two-loop and compared them to SARAH4
\cite{Staub:2013tta} where possible.
%
%This will be described in \ref{sec:kinmix}, where we also present, as an example,
%a discussion of the Standard Model supplemented by an additional $\mathrm{U}(1)_{B-L}$ symmetry.
\item \pyrate 1 supported $\mathrm{SU}(n)$ groups up to $\mathrm{SU}(6)$ 
as well as $\mathrm{U}(1)$ with the possibility to add higher  $\mathrm{SU}(n)$ gauge groups upon request. 
For each one of these groups the information on the irreducible representations (irreps) with
dimensions between $n$ and $n^2-1$ were stored in a database.
For \pyratetwo, a dedicated module, \pylie, was written to deal with all the theory aspects required for the calculation of the RGEs such that no database is required anymore. 
%This allows to deal with any irrep of $\mathrm{SU}(n)$ as well as $\mathrm{SO}(n)$ Lie algebra.
This allows us to deal with any irrep of the $\mathrm{SU}(n)$,  $\mathrm{SO}(2n)$, and
$\mathrm{SO}(2n+1)$ Lie algebras.
\item Another important new feature is that \pyrate 2 allows fields to be contracted into gauge
singlets in different ways. As a typical example, there are two possible quartic terms for a
\SU{2} triplet $\Delta$,
$$\mathrm{Tr}\left\{ \Delta^{\dagger}\Delta \Delta^{\dagger}\Delta \right\},\ \mathrm{Tr}\left\{ \Delta^{\dagger}\Delta \right\}\mathrm{Tr}\left\{ \Delta^{\dagger}\Delta \right\}\ ,$$ which can now be consistently treated in 
\pyrate 2.
\item In addition, the code has been extended in various ways:% as well be described in \ref{sec:additional}:
%\begin{itemize}
\begin{inparaenum}[(i)]
\item the two-loop expressions for the anomalous dimensions of fermion and scalar fields 
\cite{Luo:2002ti} are now available in the code,
\item it is now possible to export the coupled system of beta functions into a numerical C++ function,
\item finally, some minor bugs, collected since version 1.0.0 (currently version 1.1.9 online), have been fixed.
\end{inparaenum}
%\end{itemize}
%
\end{itemize}

The rest of this paper is organized as follows:
In \ref{sec:first_steps} we briefly summarize how to install and use \pyrate and discuss some of the new options available in \pyrate 2.
\ref{sec:pylie} introduces the new module \pylie and the model building capabilities.
Our implementation of the effects of the gauge kinetic mixing
on the two-loop RGEs of a general gauge field theory
is described in \ref{sec:kinmix}, where we also present, as an example,
a discussion of the Standard Model supplemented by an additional $\mathrm{U}(1)_{B-L}$ symmetry.
In \ref{sec:multisinglet} we describe how to handle products of fields which can be contracted
to gauge singlets in different ways, and some additional new features of \pyrate 2 are introduced 
in \ref{sec:additional}. 
Finally, in \ref{sec:conclusions} we present our conclusions.
Results for the full set of two-loop $\beta$-functions for two popular models (SM+$\mathrm{U}(1)_{B-L}$,
SM+complex triplet) have been relegated to the appendix. 
%\\
%\IS{Why do we show these results?: F: The U(1) B-L is interesting because it shows how to recover the usual result from what PyR@TE calculates (As explained below we do the calculation in a scheme where the kinetic mixing is described by an augmented gauge coupling matrix and no explicit $\xi$ parameters.}

\section{First steps with \pyrate}
\label{sec:first_steps}
%We start by briefly summarizing the use of \pyrate for the new users and refer to \cite{Lyonnet:2013dna} for additional information.

In this section we provide a brief summary of how to use \pyrate and discuss some of the new functionalities/options available in \pyrate 2. 
For additional information we refer to \cite{Lyonnet:2013dna}.

\pyrate 2 can be downloaded from the web page \url{http://pyrate.hepforge.org} where also a manual and
tutorials are provided.
To install \pyrate 2, simply open a shell and type:
\begin{Verbatim}[numbers=left,xleftmargin=20pt,formatcom=\color{cyan}]
cd $HOME
wget http://pyrate.hepforge.org/downloads/pyrate-2.0.0.tar.gz
tar xfvz pyrate-2.0.0.tar.gz
cd pyrate-2.0.0/
\end{Verbatim}

As an indication, we list the version of the various modules with which \pyratetwo was developed. However, the code is likely to work with slightly different versions of these modules with the exception of \code{SymPy}: 
\begin{itemize}
	\item \python $\geq$ 2.7.10\footnote{\pyrate was developed with \python 2.7.10 but should work with more recent/older versions with the exception of \python 3 for which it has not been tested.} \cite{Python}
	\item \code{NumPy} $\geq$ 1.7.1 \cite{dubois.hinsen.hugunin-1996-cp} and \code{SciPy} 0.12.0 \cite{scipy:2013} 
	\item \code{SymPy} $=$ 1.0.0 \cite{sympy:2013}
	\item \code{IPython} $\geq$ 4.0.1 \cite{PER-GRA:2007}
	\item \code{PyYAML} $\geq$ 3.10 \cite{PyYAML:2013}
\end{itemize}

In order to run \pyrate the user needs to provide a simple text file, the model file,
which specifies the various elements of the model under consideration 
such as the gauge group and the particle content.
%

%The various inputs such as the gauge group and particle content can be specified by the user via simple text files; the model file. 
Each of these model files contains sections such as \verb|Groups|, \verb|Fermions|, \verb|RealScalars|, \verb|CplxScalar|, \verb|Potential|, and so on. For instance, the SM gauge group structure is specified via
\begin{Verbatim}[numbers=left,xleftmargin=20pt,formatcom=\color{cyan}]
Groups: {'U1': U1, 'SU2L': SU2, 'SU3c': SU3}
\end{Verbatim}
in which a unique identifier must be used for each of the gauge group factors, (\verb|SU2L|, \verb|SU3c|, \verb|U1|). Interactions in the potential are specified in a very similar way where we distinguish \verb|QuarticTerms|, \verb|Yukawas|, \verb|ScalarMasses|, \verb|FermionMasses|, \verb|TrilinearTerms|. For instance, the Higgs quartic term of the SM Lagrangian, ${\cal L} = \tfrac{\lambda}{2} |H|^4$, would be defined by
\begin{Verbatim}[numbers=left,xleftmargin=20pt,formatcom=\color{cyan}]
QuarticTerms: {
  '\lambda': {Fields : [H,H*,H,H*], Norm : 1/2}}
  }
\end{Verbatim}

while the up-quark Yukawa term, $-{\cal L}= Y_u \bar{Q} H^c u_R + \mathrm{h.c.}$, reads
\begin{Verbatim}[numbers=left,xleftmargin=20pt,formatcom=\color{cyan}]
Yukwas: {
  'Y_{u}': {Fields: [Qbar,H*,uR], Norm: 1},
	}
\end{Verbatim}
in which \verb|H|, \verb|H*| represent the Higgs field and its charge conjugate respectively, \verb|Qbar| the adjoint left-handed quark doublet and \verb|uR| the right-handed quark singlet.

Examples of such model files are shipped with the code and can be found in the 'models' subdirectory of the
\pyrate installation.

Once the model file has been written, \pyrate is simply run by entering the following command 
(in the directory where \pyrate is installed) passing the name of the model file with the '-m' option.
For example, for the Standard Model the command would be
\begin{Verbatim}[numbers=left,xleftmargin=20pt,formatcom=\color{cyan}]
python pyR@TE.py -m models/SM.model -a
\end{Verbatim}
Note that the \verb|-a| flag specifies \pyrate to calculate the RGEs for all the terms in the potential. By default (no flag specified), \pyrate does not calculate anything. The most common flags are \verb|-gc| (gauge couplings), \verb|-yu| (Yukawas), \verb|-qc| (quartic couplings), \verb|-v| (verbose), \verb|-tl| (calculate at two-loop).
The code was designed to provide the user with maximum flexibility with regard to what to calculate. For example, one can select to calculate the $\beta$-function of only one of the terms defined in the potential or even to neglect a specific contribution to a particular RGE. Restricting the calculation to \verb|Y_{u}| is easily done by using the \verb|--Only/-onl| flag

\begin{Verbatim}[numbers=left,xleftmargin=20pt,formatcom=\color{cyan}]
python pyR@TE.py -m models/SM.model -yu -onl ['Y_{u}']
\end{Verbatim}
As another example, the following command line would calculate the RGEs of the gauge couplings, \verb|-gc|, and the quartic term, \verb|-qc|, of the SM at one-loop neglecting the contribution $A_{abcd}$\footnote{We refer the reader to~\cite{Lyonnet:2013dna} for the definition.}, \verb|--Skip/-sk ['CAabcd']|,

\begin{Verbatim}[numbers=left,xleftmargin=20pt,formatcom=\color{cyan}]
python pyR@TE.py -m SM.model -gc -qc -sk ['CAabcd'] 
\end{Verbatim}

These options are particularly useful for quick validations when developing a model or to avoid calculating terms that are time consuming and that will be neglected in the end. The full list of options can be found in \ref{tab:options}. 
The new options in version two can be found at the bottom of the table starting with the {\it Skip} command.
The {\it -ipl, -sa, -fa, -gutn, -kin} options will be discussed in \ref{sec:additional}.

%
%\newline\IS{Shouldn't we introduce the other new options (-ipl, -sa, -fa, -gutn, -kin) via exemples or
%just describing them briefly?}

\begin{table}[h!]
\caption{List of all options that can be used to control \pyratetwo. The bottom part of the table,
starting with the {\it Skip} option, lists the new options in \pyratetwo.}
\label{tab:options}
\begin{center}
\renewcommand{\arraystretch}{1.1}
\scalebox{0.83}{
\begin{tabular}{|l|l|l|}
\hline
Option& Keyword \enspace | \enspace Default & Description\\\hline\hline
\code{-\xspace-Settings/-f}& -\enspace|\enspace - &Specify the name of a {\it .settings} file.\\\hline
\code{-\xspace-Model/-m}& \code{Model\enspace|\enspace -}&Specify the name of a {\it. model} file.\\\hline
\code{-\xspace-verbose/-v}&\code{verbose\enspace|\enspace False} &Set verbose mode.\\\hline
\code{-\xspace-VerboseLevel/-vL}&\code{VerboseLevel\enspace|\enspace Critical} &Set the verbose level: {\it Info, Debug, Critical}\\\hline
\code{-\xspace-Gauge-Couplings/-gc}&\code{Gauge-Couplings\enspace|\enspace False} &Calculate the gauge couplings RGEs. \\\hline
\code{-\xspace-Quartic-Couplings/-qc}&\code{Quartic-Couplings\enspace|\enspace False} &Calculate the quartic couplings RGEs.\\\hline
\code{-\xspace-Yukawas/-yu} &\code{Yukawas\enspace|\enspace False} &Calculate the Yukawa RGEs.\\\hline
\code{-\xspace-ScalarMass/-sm}& \code{ScalarMass\enspace|\enspace False}&Calculate the scalar mass RGEs.\\\hline
\code{-\xspace-FermionMass/-fm}&\code{FermionMass\enspace|\enspace False} &Calculate the fermion mass RGEs.\\\hline
\code{-\xspace-Trilinear/-tr}&\code{Trilinear\enspace|\enspace False}&Calculate the trilinear term RGEs.\\\hline
\code{-\xspace-All-Contributions/-a}&\code{all-Contributions\enspace|\enspace False} &Calculate all the RGEs.\\\hline
\code{-\xspace-Two-Loop/-tl}&\code{Two-Loop\enspace|\enspace False} &Calculate at two-loop order.\\\hline
\code{-\xspace-Weyl/-w}&\code{Weyl\enspace|\enspace True} &The particles are Weyl spinors.\\\hline
\code{-\xspace-LogFile/-lg}&\code{LogFile\enspace|\enspace True} &Produce a log file.\\\hline
\code{-\xspace-LogLevel/-lv}&\code{LogLevel\enspace|\enspace Info} &Set the log level: {\it Info, Debug, Critical} \\\hline
\code{-\xspace-LatexFile/-tex}& \code{LatexFile\enspace|\enspace RGEsOutput.tex}&Set the name of the \LaTeX{} output file.\\ \hline
\code{-\xspace-LatexOutput/-texOut}&\code{LatexOutput\enspace|\enspace True} &Produce a \LaTeX{} output file.\\\hline
\code{-\xspace-Results/-res}&\code{Results\enspace|\enspace ./results} &Set the directory of the results\\\hline
\code{-\xspace-Pickle/-pkl}&\code{Pickle\enspace|\enspace False} &Produce a pickle output file.\\ \hline
\code{-\xspace-PickleFile/-pf}&\code{PickleFile\enspace|\enspace RGEsOutput.pickle} &Set the name of the pickle output file.\\\hline
\code{-\xspace-TotxtMathematica/-tm}&\code{ToM\enspace|\enspace False} &Produce an output to Mathematica.\\\hline
\code{-\xspace-TotxtMathFile/-tmf}&\code{ToMF\enspace|\enspace RGEsOutput.txt} &Set the name of the Mathematica output file.\\\hline
\code{-\xspace-Export/-e}&\code{Export\enspace|\enspace False}&Produce the numerical output.\\\hline
\code{-\xspace-Export-File/-ef}&\code{ExportFile\enspace|\enspace BetaFunction.py}&File in which the beta functions are written.\\\hline\hline
\code{-\xspace-Skip/-sk}&\code{Skip\enspace|\enspace -}&Set the different terms to neglect in the calculation.\\\hline
\code{-\xspace-Only/-onl}&\code{Only\enspace|\enspace -}&Set the only terms to calculate (the others skipped).\\\hline
\code{-\xspace-Pylie/-ipl}&\code{PyLie\enspace|\enspace False}&Starts the interactive PyLie mode.\\\hline
\code{-\xspace-ScalarAnomalous/-sa}&\code{ScalarAnomalous\enspace|\enspace False}&Calculate the scalar anomalous dimensions RGEs.\\\hline
\code{-\xspace-FermionAnomalous/-fa}&\code{FermionAnomalous\enspace|\enspace False}&Calculate the fermion anomalous dimensions RGEs.\\\hline
\code{-\xspace-SetGutNorm/-gutn}&\code{SetGutNorm\enspace|\enspace False}&Gut normalization: $g_1\rightarrow \sqrt{3/5}g^{\prime}$.\\\hline
\code{-\xspace-KinMix/-kin}&\code{KinMix\enspace|\enspace False}&Ignore kinetic mixing terms.\\\hline
\end{tabular}
}
\end{center}
\end{table}

\section{Lie algebra calculations within \pyratetwo: \pylie}
\label{sec:pylie}
Within \pyrate, the Clebsch-Gordan Coefficients (CGCs) were obtained from a pre-generated database of which the content was fixed\footnote{Even though several extensions of the original database have been produced to include additional irreps, e.g. \rep{15} of $\mathrm{SU}(3)$.}. In order to provide the user with more flexibility, a new \python module dubbed PyLie was developped to deal with all the group calculations. 
This includes the calculation of the Quadratic Casimir, Dynkin index, matrix representations as well as the CGCs. Therefore, \pyratetwo is able to handle any irreducible representation of $\mathrm{SU}(n)$ as well as $\mathrm{SO}(2n)$ and $\mathrm{SO}(2n+1)$. 
\pylie consists of three main classes that represent Cartan Matrices, the Lie Algebra and the permutation group 
$\mathrm{S_n}$ and is principally a \python translation of the corresponding \mathematica methods of 
SUSYNO~\cite{Fonseca:2011sy}\footnote{Note that Lie algebra related routines are also available in form of the \mathematica package LieART\protect\cite{Feger:2012bs}.}. 
In this section, we describe how to use this module through \pyrate in order to obtain the relevant information regarding the calculation of the RGEs as well as for model building.

\subsection{Interactive \pylie}

To initiate \pyrate in an interactive mode we use the flag \verb|-ipl|:
\begin{Verbatim}[numbers=left,xleftmargin=20pt,formatcom=\color{cyan}]
python pyR@TE.py -ipl
\end{Verbatim}
This will start a prompt in which the user can request any group theory information by issuing the corresponding command:
\begin{itemize}
	\item \verb|Invariants| {\it group irreps} $\Rightarrow$ CGCs of the product of the {\it irreps} of {\it group},
	\item \verb|Matrices| {\it group irrep} $\Rightarrow$ matrix representation of the generators of {\it irrep} of {\it group},
%	\IS{matrix representation of the generators of the {\it group} in the given {\it irrep},}
	\item \verb|Casimir| {\it group irrep} $\Rightarrow$ quadratic Casimir of {\it irrep} of {\it group},
	\item \verb|Dynkin| {\it group irrep} $\Rightarrow$ Dynkin index of {\it irrep} of {\it group},
	\item \verb|DimR| {\it group irrep} $\Rightarrow$ dimension of {\it irrep} of {\it group},
	\item \verb|GetDynk| {\it group dim} $\Rightarrow$ returns the Dynkin label (of the heighest weight)
	of the irrep of dimension {\it dim}\footnote{If there is more than one irrep of dimension {\it dim} an error 
	message is returned.},
%	\newline\IS{Shouldn't this be an irreducible representation? Rather use {\it group irrep} instead of {\it group dim}?}
	
	\item \verb|AdjR| {\it group} $\Rightarrow$ returns the Dynkin label (of the heighest weight)
	of the adjoint representation of {\it group},
	
	\item \verb|FondR| {\it group} $\Rightarrow$ returns the Dynkin label (of the heighest weight)
	of the fundamental representation of {\it group},
	
	\item \verb|DimAdj| {\it group} $\Rightarrow$ returns the dimension of the adjoint representation of {\it group}.
\end{itemize}

The information provided via the interactive mode is identical to what \pyrate uses to calculate the RGEs. 
Therefore, it is easy to work out the mapping between a given notation and \pyrate. 
This is particularly important for the CGCs, obtained via {\it Invariants} since there might exist several CGCs for a given combination of fields. 

\subsection{Model building with \pylie}
\label{sec:pylie_model}

For convenience, we implemented a couple of SUSYNO methods that are useful for model building which supplement the basic functions used to calculate the RGEs:

\begin{itemize}
	\item \verb|RepsUpTo| {\it group dim} $\Rightarrow$ returns all the irreps of {\it group} that have dimension lower or equal to {\it dim}. The output is a list with terms of the form (dimension, Dynkin label). 
As an example, we can check the representations of $\mathrm{SU}(3)$ with dimension lower than 21:
\begin{Verbatim}[numbers=left,xleftmargin=20pt,formatcom=\color{cyan}]
RepsUpTo SU3 21
 >> [(1, [0, 0]), (3, [1, 0]), (3, [0, 1]), (6, [0, 2]), (6, [2, 0]),
  (8, [1, 1]), (10, [0, 3]), (10, [3, 0]), (15, [2, 1]), (15, [1, 2]),
  (15, [4, 0]), (15, [0, 4]), (21, [0, 5]), (21, [5, 0])]
\end{Verbatim}
which as expected contains the \rep{1}, \rep{3}, \rep{6}, \rep{10}, \rep{15}, \rep{21} and their respective conjugated counterparts as well as the \rep{15'}. 
\item \verb|ReduceProduct| {\it group irreps} $\Rightarrow$ calculates the 
Clebsch-Gordan decomposition of the tensor
product of the {\it irreps} of {\it group}. The output is a list of terms of the form ((Dynkin label, multiplicity), dimension). For instance, the result for $\rep{15}\times\crep{15}\times\rep{15}\times\crep{15}$ reads:
\begin{Verbatim}[numbers=left,xleftmargin=20pt,formatcom=\color{cyan}]
ReduceProduct SU3 [[1,2],[2,1],[1,2],[2,1]]
 >> [(([0, 0], 14), 1), (([1, 1], 72), 8), (([3, 0], 59), 10),
 (([0, 3], 59), 10), (([2, 2], 121), 27), (([0, 6], 28), 28),
 (([6, 0], 28), 28), (([1, 4], 88), 35), (([4, 1], 88), 35),
 (([0, 9], 2), 55), (([9, 0], 2), 55), (([3, 3], 104), 64),
 (([7, 1], 18), 80), (([1, 7], 18), 80), (([2, 5], 60), 81),
 (([5, 2], 60), 81), (([4, 4], 49), 125), (([3, 6], 21), 154),
 (([6, 3], 21), 154), (([8, 2], 4), 162), (([2, 8], 4), 162),
 (([5, 5], 12), 216), (([4, 7], 3), 260), (([7, 4], 3), 260),
 (([6, 6], 1), 343)]
\end{Verbatim}
As can be seen, the decomposition contains 14 singlets.
\item \verb|PermutationOfSymmetryInvs| {\it group irreps} $\Rightarrow$ returns all the information on the permutation symmetries of the invariants. Its output contains two parts: 
\begin{inparaenum}[(i)]
\item It first lists the groups of indices that mix together. 
Let $k$ be the number of groups and $n_i$ ($i=1,\ldots,k)$ the number of indices in the $i$th group.
\item This is followed by a list of the irreps of 
$\mathrm{S_{n_1}}\times\mathrm{S_{n_2}}\times\ldots\times \mathrm{S_{n_k}}$
under which the invariants transform.
\end{inparaenum}~Let's exemplify this with the above product $\rep{15}\times\crep{15}\times\rep{15}\times\crep{15}$ which 
contains 14 invariants that can be decomposed as:
\begin{Verbatim}[numbers=left,xleftmargin=20pt,formatcom=\color{cyan}]
PermutationSymmetryOfInvs SU3 [[1,2],[2,1],[1,2],[2,1]]
 >>[[[1, 3], [2, 4]],
 [[[(1, 1), (1, 1)], 5], [[(2), (2)], 5],
 [[(1, 1), (2)], 2], [[(2), (1, 1)], 2]]]
\end{Verbatim}
The first item, \verb|[[1, 3], [2, 4]]| tells us that the first and third fields are identical as well as the second and fourth
i.e., $k=2, n_1 = 2, n_2 = 2$. 
Then it shows that the 14 invariants decompose into 4 irreps of $\mathrm{S_2}\times\mathrm{S_2}$:
\begin{inparaenum}[(a)]
\item The \rep{(\{1,1\},\{1,1\})} with multiplicity 5 which is the one fully anti-symmetric under the exchange of either the 1$^{\mathrm{st}}$ and $3^{\mathrm{rd}}$ or the 2$^{\mathrm{nd}}$ and $4^{\mathrm{th}}$ indices. 
\item  The \rep{(\{2\},\{2\})} again with multiplicity 5
 which is fully symmetric under the exchange of the indices among the first (second) group respectively. 
\item
Finally, with multiplicity 2, the representations \rep{(\{1,1\},\{2\})} and \rep{(\{2\},\{1,1\})} with mixed symmetry
properties.
\end{inparaenum}
\item \verb|SnIrrepDim| {\it irrep} $\Rightarrow$ returns the dimension of an $\mathrm{S_n}$ irrep. This allows one to check for instance that indeed the \rep{\{1,1\}} and \rep{\{2\}} of $\mathrm{S_2}$ are of dim 1.
\begin{Verbatim}[numbers=left,xleftmargin=20pt,formatcom=\color{cyan}]
SnIrrepDim [1,1]
 >> 1
SnIrrepDim [2]
 >> 1
\end{Verbatim}

\end{itemize}

Finally, note that \pyratetwo still has a database which is being updated each time \pyrate is run either to calculate some RGEs or querying results through \pylie. This allows us to greatly improve the execution time when large representations are used. The interactive mode is exited via the command \verb|exit| or \verb|quit|.

\section{Kinetic Mixing at two-loop}
\label{sec:kinmix}

Multiple $\mathrm{U}(1)$ group factors are present in many BSM theories
and an appropriate treatment of the mixing of the gauge kinetic terms is
indispensable for studying the scale dependence of these theories. 
Indeed, it has been shown that the effects of kinetic mixing (in the RGEs) can
be important \cite{Rizzo:1998ut}.
In this section, we summarize the modifications of the RGEs in general gauge theories
following the approach in \cite{Fonseca:2013jra} and discuss how these modifications
are implemented in \pyratetwo.
We illustrate the effects of kinetic mixing for the SM supplemented by an additional $\mathrm{U}(1)_{B-L}$
symmetry.

\subsection{Kinetic mixing in a general gauge field theory}

The impact of kinetic mixing on the two-loop RGEs of the dimensionless parameters of general gauge field theories has been calculated in \cite{Luo:2002iq}. Recently, a new approach has been proposed and applied first to supersymmetic (SUSY) and then to non-SUSY theories~\cite{Fonseca:2013bua,Fonseca:2013jra}. 
%
%The approach of~\cite{Fonseca:2013bua,Fonseca:2013jra} 
This approach is particularly well suited for implementation 
in a computer code
as there is no need for new parameters\footnote{However, the gauge couplings must be promoted to a non-diagonal matrix of effective gauge couplings.} compared to the case without kinetic mixing and the required set of modifications to be applied to the RGEs of Machacek and Vaughn is minimal. We now review the main features of this approach.

With $n$ $\mathrm{U}(1)$ group factors, the field strength tensors related to different $\mathrm{U}(1)$ can mix as expressed by the term in the Lagrangian
\begin{equation}
	\mathcal{L}_{\mathrm{kin.}} \ni -\frac{1}{4} F^{T}_{\mu\nu}\xi F^{\mu\nu}\, .
	\label{eq:kinlagrangian}
\end{equation}
$F^{\mu\nu}$ is a vector of $n$ tensor fields corresponding to the $\mathrm{U}(1)$ group factors. Correspondingly, $\xi$ is a $n\times n$ real symmetric matrix introducing $\frac{1}{2}n(n-1)$ extra dynamical parameters
(in addition to the $n$ diagonal parameters which are present also if the kinetic mixing is neglected). 
The approach followed by~\cite{Luo:2002iq} is to calculate the beta functions for each one of these new parameters as well as the modifications induced to the RGEs of the other parameters.

An equivalent approach~\cite{Fonseca:2013bua,Fonseca:2013jra} is to trade these $\frac{1}{2}n(n-1)$ parameters with effective gauge couplings. These new gauge couplings populate the off diagonal terms of an extended gauge coupling matrix
\begin{equation}
	\label{eq:gaugematrix}
G\equiv\tilde{G}\xi^{-1/2}\, ,
\end{equation}
with $\tilde{G}$ the original diagonal gauge coupling matrix. This is easily seen by performing a redefinition of the vector fields $V^{\prime} = \xi^{1/2}V$, and we refer the reader to~\cite{Fonseca:2013bua,Fonseca:2013jra} for more details. The replacement rules are then derived by substituting the polynomials involving the gauge couplings by the adequate matrix structure.

There are a couple of key points/features that make this approach interesting: 
\begin{inparaenum}[(i)]
\item there is no new parameter for which one needs to calculate the beta function, 
\item the list of replacement rules is short, and
\item the rules are simpler to apply as they take a simple matrix form.
\end{inparaenum}
	
As an illustration, we show here the results for the replacement rules involving the scalar generators, 
$\Theta_A^{\alpha}$. 
We denote $C_A(R)$($S_A(R)$) the quadratic Casimir (Dynkin index) of the representation $R$ 
%\IS{of the representation $R$ (since the index 'i' doesn't appear)?}
of  the gauge group $A$ and $C(\mathcal{G})$ the quadratic Casimir of the adjoint representation of 
the group $\mathcal{G}$.	
Correcting a typo in Eq.~(23) of \cite{Fonseca:2013bua} the replacement rules read
	\begin{eqnarray}
		g^2\left( \Theta_A^{\alpha} \right)_{ac} \left( \Theta_{A}^{\alpha} \right)_{bd} &\rightarrow& \tilde{\Lambda}_{ab,cd}\, ,\\
		g^4 \left\{\Theta^{\alpha}, \Theta^{\beta}\right\}_{ab} \left\{ \Theta^{\alpha},\Theta^{\beta} \right\}_{cd} &\rightarrow& 2 \sum_{e,f} \left( \tilde{\Lambda}_{ac,ef}\tilde{\Lambda}_{ef,bd} +\tilde{\Lambda}_{af,ed}\tilde{\Lambda}_{ec,bf} \right)\, ,\label{eq:thematrep}\\
		g^6 C(\mathcal{G})\left\{\Theta^{\alpha}, \Theta^{\beta}\right\}_{ab} \left\{ \Theta^{\alpha},\Theta^{\beta} \right\}_{cd} &\rightarrow& 2 \sum_{e,f} \left( \tilde{\Lambda}^{\mathcal{G}}_{ac,ef}\tilde{\Lambda}_{ef,bd} +\tilde{\Lambda}^{\mathcal{G}}_{af,ed}\tilde{\Lambda}_{ec,bf} \right)\, ,\\
		g^6 S(R)\left\{\Theta^{\alpha}, \Theta^{\beta}\right\}_{ab} \left\{ \Theta^{\alpha},\Theta^{\beta} \right\}_{cd} &\rightarrow& 2 \sum_{e,f} \left( \tilde{\Lambda}^{S}_{ac,ef}\tilde{\Lambda}_{ef,bd} +\tilde{\Lambda}^{S}_{af,ed}\tilde{\Lambda}_{ec,bf} \right)\, ,\\
		g^{6}\sum_i C(i) \left\{ \Theta^{\alpha},\Theta^{\beta} \right\}_{ab} \left\{ \Theta^{\alpha},\Theta^{\beta} \right\}_{cd} &\rightarrow&2\sum_{e,f}\sum_{\mathrm{free}\ i}\left[g_B^2 C_{B}(i) + \left( W_{i}^{S} \right)^{T}W_{i}^{S} \right]\left( \tilde{\Lambda}_{ac,ef}\tilde{\Lambda}_{ef,bd} +\tilde{\Lambda}_{af,ed}\tilde{\Lambda}_{ec,bf} \right),
		\label{reprules}
	\end{eqnarray}
where $i$ runs over the free indices, e.g. $i=a,b,c,d$ for the quartic terms. 
The following short hand notation has been defined
\begin{eqnarray}
	\tilde{\Lambda}_{ab,cd} &\equiv& \sum_{A}g_{A}^{2}\left( \Theta_{A}^{\alpha} \right)_{ac}\left( \Theta_{A}^{\alpha} \right)_{bd} + \tilde{\delta}_{ac}\tilde{\delta}_{bd} \left( W^{S}_{a} \right)^{T}W^{S}_b\, ,\\
	\tilde{\Lambda}^{S}_{ab,cd} &\equiv& \sum_{A}g_A^{4} S_{A}(R) \left( \Theta_{A}^{\alpha} \right)_{ac}\left( \Theta_{A}^{\alpha} \right)_{bd} + \tilde{\delta}_{ac}\tilde{\delta}_{bd}\sum_p \left( W^{S}_{a} \right)^{T}W_p^{R}\left( W^{R}_p \right)^{T}W_{b}^{S}\, ,\\
	\tilde{\Lambda}^{\mathcal{G}}_{ab,cd}&\equiv& \sum_{A}g_A^{4} C(\mathcal{G}_A)\left( \Theta_{A}^{\alpha} \right)_{ac}\left( \Theta_{A}^{\alpha} \right)_{bd}\, .
\end{eqnarray}
Finally, in the above equation we made use of $W^{R}_i \equiv G^T Q_i^{R}$ with $Q_i^R$ 
the column vector of charges of a given field and the antisymmetric tensor $\tilde{\delta}_{ab}$ equal to the imaginary unit $i$ if $`a`$ and $`b`$ are the real and imaginary components of an eigenstate of the $\mathrm{U}(1)$ gauge interactions, respectively, and zero otherwise.

%It is important to note that the modifications to the various beta functions can be factorized such that one can write
%\begin{equation}
%	\beta_{i} = \beta_{\mathrm{no\ kin.}} + \beta_{\mathrm{kin.}}\, ,
%\end{equation}
%and the implementation can be carried out without modifying the existing beta functions, hence limiting the possible mistakes.

\subsection{Kinetic Mixing in \pyrate}

We implemented the complete set of replacement rules~\cite{Fonseca:2013bua} in \pyratetwo such that all the effects of the kinetic mixing are consistently taken into account at two-loop. 
These modifications are automatically switched on when several $\mathrm{U}(1)$ group factors are present in the model file. To allow the user to ignore these terms we implemented the \codeq{-\xspace-KinMix/-kin} switch.

Once the RGEs have been computed, \pyrate exports the results as usual into the different formats specified by the user. Note however, that the non-diagonal entries of the gauge matrix defined in~\ref{eq:gaugematrix} are denoted by $g_{Uij}$. The difference in speed between taking into account the kinetic mixing terms and neglecting them has been found to be minimal as for most of the terms the additional contributions are obtained by matrix multiplications.

\subsection{SM-$U(1)_{B-L}$ example}
As an example, we apply our implementation to the SM-$\mathrm{U}(1)_{B-L}$ where the SM gauge group
$\mathrm{SU}(3)_c\times\mathrm{SU}(2)_L\times\mathrm{U}(1)_Y$ is supplemented by an 
additional $\mathrm{U}(1)_{B-L}$ group factor with $B-L$ charge where $B$ and $L$ are the 
baryon and lepton number, respectively. 
In addition to the SM particle spectrum we also consider three right-handed neutrinos and an extra complex scalar $\chi\sim(1,1,0,-2)$. This setup has been extensively studied in the literature and in particular the stability of the electroweak vacuum has been analysed~\cite{Coriano:2015sea,Basso:2010jm, Datta:2013mta}. 
	
As explained above, we work with the full $2\times 2$ gauge coupling matrix 
\begin{equation}
	G\equiv\left(\begin{array}{cc}g_{YY}&g_{YB}\\g_{BY}&g_{BB}\end{array}\right)\, .
\end{equation}
	In this basis, the covariant derivative for a field $\phi$ charged under the two Abelian group factors reads 
\begin{equation}
	D_{\mu} = \partial_{\mu} - i Q^T_\phi G A_{\mu}\, ,
\end{equation}
with the column vector of vector bosons $A_{\mu}=\left(\begin{array}{c}A^{Y}_{\mu}\\A^{B-L}_{\mu}\end{array}\right)$. 

Because the matrix $\xi$ contains $\frac{1}{2}n(n-1)=1$ mixing parameter, by extension the matrix $G$ only has three independent parameters. Therefore, it is usual to work in the triangular basis in which these degrees of freedom are apparent. The two basis are related by an orthogonal rotation $G^{\prime} = GO^{T}_{R} = \left(\begin{array}{cc}g&\tilde{g}\\0&g^{\prime}\end{array}\right)$. Hence, to obtain the RGEs of $(g,\tilde{g},g^{\prime})$ from $(g_{YY},g_{BY},g_{YB},g_{BB})$ one must 
\begin{enumerate}[(i)]
	\item first compute the derivative of the product $GO_{R}^{T}$: $d(GO_{R}^T)/dt\equiv\left(\begin{matrix}\beta_{g}&\beta_{\tilde{g}}\\0&\beta_{g'}\end{matrix}\right)=(dG/dt) O_{R}^T + G(dO_{R}^{T}/dt)$,
	\item substitute the derivatives of the couplings, $dg_{YY}/dt\equiv\beta_{g_{YY}},dg_{YB}/dt\equiv\beta_{g_{YB}},dg_{BY}/dt\equiv\beta_{g_{YB}},dg_{BB}/dt\equiv\beta_{g_{BB}}$,  by their expressions as given by \pyrate 
	and finally,
\item express the result in terms of $g,\tilde{g},g^{\prime}$ using $G=G^{\prime}O_R$.
\end{enumerate}
	
Recently, the two-loop RGEs in this basis have been obtained in~\cite{Coriano:2015sea} in the limit $Y_{\nu}\simeq 0$ against which we validated our results.
In addition, we provide the beta functions for the scalar mass terms. 
For the discussion we show only the one-loop expression for all the couplings other than the gauge couplings 
and relegate the two-loop expressions to~\ref{app:twobl}.

With two scalar fields, the effective potential can be written in the following form~\cite{Coriano:2015sea}
	\begin{equation}
		\mathcal{V}=\lambda_1 H^{\dagger} H H^{\dagger} H+ \lambda_2 \chi^{\dagger}\chi \chi^{\dagger}\chi+\lambda_3 (H^{\dagger}H) (\chi^{\dagger}\chi)+\mu H^{\dagger}H +\mu_\chi \chi^{\dagger}\chi\, , 
	\end{equation}
	and additional Yukawa terms have to be added to the Yukawa sector of the SM Lagrangian
	\begin{equation}
		-\mathcal{L}^{Yuk} = -\mathcal{L}^{Yuk}_{SM} + Y_{\nu}\bar{L}H^{c}\nu_{R} + Y_{N}\overline{\nu_R^c}\nu_R \chi 
	+ h.c. \, .
	\end{equation}
Note that this model implements a type-I seesaw scenario for the three light neutrinos~\cite{Coriano:2015sea} 
with a Majorana mass term, $M\overline{\nu_R^c}\nu_R$, dynamically generated by the 
vacuum expectation value
%vev 
of the $\chi$ field.
For the sake of comparison with results in the literature we present results in which we neglect $Y_{d}$ and $Y_{e}$ and retain only the Yukawa coupling of the top quark, $Y_t$. However, we keep the full dependence on $Y_{\nu}$. Finally, we assume the Yukawa couplings of the heavy neutrinos to be universal, i.e., $Y_N^{ij}=\delta^{ij}Y_{N}$. 
However, the full results are easily obtained running \pyratetwo. In addition, a detailed study of the numerical impact of the kinetic mixing in this model is presented in~\cite{Lyonnet:2016}. Note that all the results for an arbitrary coupling $x$ are given in the form
\begin{equation}
	\beta_{x} = \frac{1}{(4\pi)^2}\beta_x^{(1)} + \frac{1}{(4\pi)^{4}}\beta_x^{(2)}\, .
\end{equation}

	\subsubsection*{Gauge Couplings}

At one-loop the beta functions for the non-Abelian gauge couplings ($g_2(\mathrm{SU}(2)), g_3 (\mathrm{SU}(3))$ 
do not get modified by kinetic mixing and read
\begin{eqnarray}
	\beta^{(1)}_{g_2} \equiv \frac{d}{dt}g_2 &=& -\frac{19}{6}g_2^3\, ,\\
	\beta^{(1)}_{g_3} \equiv \frac{d}{dt}g_3 &=& -7 g_3^3\, .
\end{eqnarray}
The Abelian part involves the three couplings $g^{\prime},g,\tilde{g}$ and their products
\begin{eqnarray}
	\beta^{(1)}_g \equiv \frac{d}{dt}g &=& \frac{41 g^3}{6}\, ,\\
	\beta^{(1)}_{\tilde{g}} \equiv \frac{d}{dt}\tilde{g} &=& \frac{41 g^2 \tilde{g}}{3}+\frac{32}{3} \tilde{g}^2 g'+12 \tilde{g} (g^{\prime})^2+\frac{41 \tilde{g}^3}{6}+\frac{32 g^2 g^{\prime}}{3}\, ,\\
	\beta^{(1)}_{g^{\prime}} \equiv \frac{d}{dt} g^{\prime} &=& \frac{32}{3} \tilde{g} (g^{\prime})^2+\frac{41}{6} \tilde{g}^2 g^\prime+12 g'^3\, .
\end{eqnarray}
At two-loop, the non-Abelian beta functions get an additional contribution coming from Eq.\ (16) of 
\cite{Fonseca:2013jra} leading to
\begin{eqnarray}
	\beta^{(2)}_{g_2} &=& 4 g_2^3 \tilde{g} g'+\frac{3}{2} g_2^3 \tilde{g}^2+\frac{3}{2} g^2 g_2^3+4 g_2^3 \left(g'\right)^2-\frac{3}{2} g_2^3 Y_t^2+\frac{35 g_2^5}{6}+12 g_3^2 g_2^3\, ,\label{eq:g2bl}\\
	\beta^{(2)}_{g_3} &=&		\frac{4}{3} g_3^3 \tilde{g} g'+\frac{11}{6} g_3^3 \tilde{g}^2+\frac{11}{6} g^2 g_3^3+\frac{4}{3} g_3^3 \left(g'\right)^2-2 g_3^3 Y_t^2-26 g_3^5+\frac{9}{2} g_2^2 g_3^3\, .\label{eq:g3bl}
\end{eqnarray}
Note that \ref{eq:g2bl} and \ref{eq:g3bl} agree with~\cite{Coriano:2015sea}. The beta functions for the Abelian gauge couplings read
\begin{eqnarray}
	\beta^{(2)}_{g} &=& \frac{1}{18} g^3 \left(328 \tilde{g} g'+199 \tilde{g}^2+199 g^2+184 \left(g'\right)^2+81 g_2^2+264 g_3^2-51 Y_t^2-9 Y_{\nu }^2\right)\, ,\\ 
	\beta^{(2)}_{\tilde{g}} &=&\frac{1}{18} \left(656 \tilde{g}^4 g'+3 \tilde{g}^3 \left(199 g^2+368 \left(g'\right)^2+27 g_2^2+88 g_3^2-17 Y_t^2-3 Y_{\nu }^2\right)\right.\nonumber\\
	&\phantom{.}&\left.+4 \tilde{g}^2 g' \left(328 g^2+224 \left(g'\right)^2+54 g_2^2+48 g_3^2-15 Y_t^2-9 Y_{\nu }^2\right)\right.\nonumber\\
	&\phantom{.}&+2 \tilde{g} \left(199 g^4-51 g^2 Y_t^2-9 g^2 Y_{\nu }^2-12 \left(g'\right)^2 Y_t^2-36 \left(g'\right)^2 Y_{\nu }^2-36 Y_N^2 \left(g'\right)^2+800 \left(g'\right)^4\right.\nonumber\\
	&\phantom{.}&\left.+644 g^2 \left(g'\right)^2+27 g_2^2 \left(3 g^2+4 \left(g'\right)^2\right)+24 g_3^2 \left(11 g^2+4 \left(g'\right)^2\right)\right)+199 \tilde{g}^5\nonumber\\
	&\phantom{.}&\left.+4 g^2 g' \left(82 g^2+112 \left(g'\right)^2+54 g_2^2+48 g_3^2-15 Y_t^2-9 Y_{\nu }^2\right)\right)\, , \\
	\beta^{(2)}_{g^{\prime}} &=&  \frac{1}{18} g' \left(656 \tilde{g}^3 g'+\tilde{g}^2 \left(199 g^2+1104 \left(g'\right)^2+81 g_2^2+264 g_3^2-51 Y_t^2-9 Y_{\nu }^2\right)\right.\nonumber\\
	&\phantom{.}&+4 \tilde{g} g' \left(82 g^2+224 \left(g'\right)^2+54 g_2^2+48 g_3^2-15 Y_t^2-9 Y_{\nu }^2\right)+199 \tilde{g}^4\nonumber\\
	&\phantom{.}&\left.+8 \left(g'\right)^2 \left(23 g^2+200 \left(g'\right)^2+27 g_2^2+24 g_3^2-3 Y_t^2-9 Y_{\nu }^2-9 Y_N^2\right)\right)\, .
\end{eqnarray}
where, after detailed comparison, we now find complete agreement with~\cite{Coriano:2015sea} for the terms involving $Y_t$ and $Y_N$\footnote{Note that the authors of~\cite{Coriano:2015sea} updated their original result accordingly in a new version of their article.}. 
Note that our results also agree with the ones from SARAH~\cite{Staub:2013tta}.

\subsubsection*{Yukawa couplings}
The one-loop beta functions for the Yukawa couplings are given by
\begin{eqnarray}
	\beta^{(1)}_{Y_t} &=& \frac{1}{12} Y_t \left(-20 \tilde{g} g'-17 \tilde{g}^2-17 g^2-8 \left(g'\right)^2-27 g_2^2-96 g_3^2+54 Y_t^2+12 \text{Tr}\left( Y_{\nu}Y_{\nu}^{\dagger} \right)\right)\label{eq:ytbl}\, ,
	\\
	\beta^{(1)}_{Y_{N}} &=& Y_N \left(-6 \left(g'\right)^2+10 Y_N^2+Y_{\nu }^{\dagger}Y_{\nu} + Y_{\nu}^{t}Y_{\nu}^{*}\right) \label{eq:ynbl}\, ,
	\\
	\beta^{(1)}_{Y_\nu} &=&\frac{1}{4} Y_{\nu } \left(-12 \tilde{g} g'-3 \tilde{g}^2-3 g^2-24 \left(g'\right)^2-9 g_2^2+8 Y_N^2+12 Y_t^2+ 4 \text{Tr}\left( Y_{\nu }Y_{\nu}^{\dagger}\right)\right) + \frac{3}{2}Y_{\nu}Y_{\nu}^{\dagger}Y_{\nu}\, .
	\label{eq:ynubl}
\end{eqnarray}
Our expressions in \ref{eq:ytbl}, \ref{eq:ynbl} and \ref{eq:ynubl} agree with the results found 
in~\cite{Coriano:2015sea, Basso:2010jm}\footnote{The calculation of \cite{Basso:2010jm} is at one-loop and we therefore perform the comparison with their results in this limit.} and SARAH. 
\subsubsection*{Quartic couplings}
Our SM-$\mathrm{U}(1)_{B-L}$ model has three quartic couplings for which we find the following
beta functions:
\begin{eqnarray}
	\beta^{(1)}_{\lambda_1}&=&\frac{3}{4} \tilde{g}^2 \left(g^2+g_2^2-4 \lambda _1\right)+\frac{3 \tilde{g}^4}{8}+\frac{3 g^4}{8}-3 g^2 \lambda _1+\frac{3}{4} g_2^2 \left(g^2-12 \lambda _1\right)\\
	&\phantom{.}&+\frac{9 g_2^4}{8}+24 \lambda _1^2+\lambda _3^2+12 \lambda _1 Y_t^2-6 Y_t^4+4 \lambda _1 \text{Tr}\left(Y_{\nu }Y_{\nu }^{\dagger}\right)-2 \text{Tr}\left(Y_{\nu }Y_{\nu }^{\dagger}Y_{\nu }Y_{\nu }^{\dagger}\right)\, ,\\
	\beta^{(1)}_{\lambda_2}&=& 24 \lambda _2 \left(Y_N^2-2 \left(g'\right)^2\right)+96 \left(g'\right)^4+20 \lambda _2^2+2 \lambda _3^2-48 Y_N^4\, ,\\
	\beta^{(1)}_{\lambda_3} &=& \frac{1}{2} \left(24 \tilde{g}^2 \left(g'\right)^2+\lambda _3 \left(-3 \tilde{g}^2-3 g^2-48 \left(g'\right)^2-9 g_2^2+24 \lambda _1+16 \lambda _2+12 Y_t^2+24 Y_N^2\right)\right.\\
	&\phantom{.}&\left.+8 \lambda _3^2+\text{Tr}\left( Y_{\nu}Y_{\nu}^{\dagger} \right)\left(4 \lambda _3-32 Y_N^2\right)\right)\, .
\end{eqnarray}
Again we have compared our results with the literature where possible. 
In the limit of no kinetic mixing, our results agree with SARAH. Taking the kinetic mixing into account, 
there are some differences in the terms of order ${\cal O}(g^4)$ involving at least one Abelian coupling. 
This is easily traced back to the fact that \ref{eq:thematrep} has not been implemented
in SARAH. We find perfect agreement with~\cite{Coriano:2015sea} limiting our results to their theoretical assumptions stated above.

\subsubsection*{Scalar mass terms}
Finally, we provide our results for the scalar mass beta functions at one-loop:
\begin{eqnarray}
	\beta^{(1)}_{\mu}&=& \frac{1}{2} \left(-3 \mu  \tilde{g}^2-3 g^2 \mu -9 g_2^2 \mu +4 \lambda _3 \mu _{\chi }+24 \lambda _1 \mu +12 \mu  Y_t^2+4 \mu  \text{Tr}\left( Y_{\nu}Y_{\nu}^{\dagger} \right)\right)\, ,\\
	\beta^{(1)}_{\mu_\chi} &=&4 \left(\mu _{\chi } \left(-6 \left(g'\right)^2+2 \lambda _2+3 Y_N^2\right)+\lambda _3 \mu \right)\, .
\end{eqnarray}
The two-loop expressions for the Yukawa couplings, the quartic couplings, and the
scalar mass terms can be found in \ref{app:twobl}.
It is important to note that perfect agreement with~\cite{Coriano:2015sea} was obtained for the provided parameters. In addition to these parameters, we also provide the RGEs for $Y_{\nu},\mu$ and $\mu_{\chi}$.

\section{Multiple Invariants}
\label{sec:multisinglet}
	
This section describes another addition to the code made recently~\cite{Lyonnet:2015dna} and merged to \pyratetwo: the possibility to deal with multiple gauge singlets. 
The main motivation is to be able to automatically calculate the RGEs 
in situations where products of fields are present that allow for multiple ways to contract them into gauge singlets (see \ref{sec:pylie_model}). 
Indeed, as already mentioned above, each term in the Lagrangian is specified in the model file by a set of fields. This makes it impossible to distinguish multiple singlets that could result from the contraction of the same set of fields.
In \pyrate 1, the chosen combination was the one in which the quartic term can be factorized in
\begin{equation}
	\underbrace{(\rep{a}\otimes\rep{b})}_{\rep{1}}\otimes\underbrace{(\rep{c}\otimes\rep{d})}_{\rep{1}}\, ,
\end{equation}
where $\rep{a},\ \rep{b},\ \rep{c},\ \rep{d}$ are the representations of the scalar fields involved. 
In general, assuming a set of $k$ fields  $\phi_i$, where
each $\phi_i$ belongs to a $D_i$-dimensional irrep of an $\mathrm{SU}(n)$ gauge group, 
%with dimensions $D_i$ under an $\mathrm{SU}(n)$ gauge group, 
we will denote the Clebsch-Gordan coefficients that give the contraction of indices to an invariant combination as 
$\mathcal{C}$, i.e.
\begin{equation}
	 {\cal C}_{i_1 i_2 \dots i_k}   \phi_{i_1} \phi_{i_2} \dots \phi_{i_k}\, ,
 \end{equation}
 does not transform under $\mathrm{SU}(n)$. Here, the $i_x$ ($x=1, \ldots, k$) are the charge indices 
 with respect to the gauge group. This section describes how to implement terms that have two or more such invariants (${\cal C}^{1},\ \mathcal{C}^2, \ldots$) in \pyratetwo.	
 
 A typical example is that of a complex triplet of $\mathrm{SU}(2)$. %, $T$. 
Writing the triplet as a two-by two-matrix $\Delta$, two quartic invariants 
can be formed which are usually written in the literature as
\begin{eqnarray}
	&\mathrm{Tr}(\Delta^{\dagger}\Delta\Delta^{\dagger}\Delta)\, ,\label{eq:exsu2_1}\\
	&\mathrm{Tr}(\Delta^{\dagger}\Delta)\mathrm{Tr}(\Delta^{\dagger}\Delta)\, .\label{eq:exsu2_2}
\end{eqnarray}

When there are several invariant contractions in a product of representations, $\{\mathcal{C}^1,\ \mathcal{C}^2,\ \dots,\ \mathcal{C}^m\}$, any linear combination of the $\mathcal{C}^i$ is also an invariant. More specifically, we can trade the set $\{\mathcal{C}^1,\ \mathcal{C}^2,\ \dots,\ \mathcal{C}^m\}$ for $\{\sum_{j=1}^mA_{ij}\mathcal{C}^{j}\}_{i=1,\dots,m}$ where $A_{ij}$ is an arbitrary non-singular $m\times m$ matrix. 
Even if one requires a specific normalization for the CGCs, one is left with an immense number of possibilities. It is therefore reasonable to use this freedom and write the invariants in an advantageous form. 
Assuming we have the product of $k$ identical fields, the invariants will be in representations of $S_k$ which one would like to be irreducible. Following~\cite{Fonseca:2011sy} the strategy of \pylie is to segregate the invariants in terms of irreps of $S_k$. 

For example, it is common knowledge that the invariant of the product of two doublets of $\mathrm{SU}(2)$ is in the anti-symmetric irrep of $S_{2}$. Going forward, let us now consider the product $\rep{3}\otimes \rep{3} \otimes \crep{3}\otimes \crep{3}$ of four representations of $\mathrm{SU}(2)$. 
In the general case where all four fields are different there exist three different invariants in $S_2\times S_2$ that are segregated in the following way:
	\begin{enumerate}[(i)]
		\item two are in the $(\{2\},\{2\})$ of $S_2\times S_2$, i.e. they are completely symmetric under the exchange of the first two fields or the last two,
		\item and one is in the $(\{1,1\},\{1,1\})$ which is anti-symmetric under the exchange of the first two or last two fields.
	\end{enumerate}	

To distinguish the various invariants in the model file we introduce a new keyword, {\it CGCs}.

\subsection{The {\it CGCs} keyword}

The user has the possibility to specify % \IS{(has the possibility to specify or is this mandatory?)}
which one of the CGCs to use for each of the Lagrangian terms via a new keyword, \verb|CGCs|. As an example, assuming that the invariant defined by \ref{eq:exsu2_1} corresponds to the second CGC returned by \pyrate, $\mathcal{C}^{2}$, one would specify this term via
\begin{Verbatim}[numbers=left,xleftmargin=20pt,formatcom=\color{cyan}]
QuarticTerms: {
 '\lambda': {Fields : [T*,T,T*,T], Norm : 1, CGCs: {SU2L: [2]}}
}
\end{Verbatim}
Note that the argument of the CGCs keyword must be a list to accommodate the possibility that \ref{eq:exsu2_1} is given by a linear combination of the two invariants of \pyrate. 
An example is given below in \ref{sec:SM+triplet}.
If the \verb|CGCs| keyword is not specified then \pyratetwo returns the first one.

\subsection{A toy model: SM+complex triplet}
\label{sec:SM+triplet}

As an example, let us consider the SM extended by a triplet of $\mathrm{SU}(2)_L$ of complex scalars, 
$T\equiv(\Delta^{++},\Delta^+,\Delta^0)$. With the two scalar fields $H\sim (2,1/2)$ and $T\sim(3,1)$ 
(transformed into a $2\times 2$ matrix $\Delta = \sigma_i T_i$ where $\sigma_{i=1,2,3}$ are the Pauli matrices)
we write the following potential
\begin{eqnarray}
	\mathcal{V} =& \lambda_1 H^\dagger H H^{\dagger} H + \lambda_{2}\mathrm{Tr}(\Delta^{\dagger}\Delta)H^{\dagger}H +\lambda_3 H^{\dagger}\Delta \Delta^{\dagger}H + \lambda_{\Delta_1}\mathrm{Tr}(\Delta^{\dagger}\Delta)\mathrm{Tr}(\Delta^{\dagger}\Delta)+\lambda_{\Delta_2} \mathrm{Tr}(\Delta^{\dagger}\Delta\Delta^{\dagger}\Delta)\, .
	\label{eq:potentialToyTII}
\end{eqnarray}
As explained in~\ref{sec:pylie}, one obtains the list of CGCs implemented in \pyrate for the contraction of four arbitrary triplet fields of $\mathrm{SU}(2)$ ($\rep{a},\ \rep{b},\ \rep{c},\ \rep{d}$) by using \pylie.
For instance,
\begin{Verbatim}[numbers=left,xleftmargin=20pt,formatcom=\color{cyan}]
Invariants SU2 [[2,True],[2],[2,True],[2]]
\end{Verbatim}
leads to three invariants\footnote{Note that $`[2]`$ is the Dynkin label of the \rep{3} of \SU{2}.}, $\mathcal{C}^{1,2,3}$. Once the substitutions $\rep{a},\rep{c}\rightarrow T^{\dagger}$ 
and $\rep{b},\rep{d}\rightarrow T$ have been performed, $\mathcal{C}^{1}$ vanishes and one is left with two invariants.
Working out the details of the matching between these invariants and the ones in \ref{eq:exsu2_1}, \ref{eq:exsu2_2} one arrives at the following relations for the CGCs of the $\lambda_{\Delta_1}$ and $\lambda_{\Delta_2}$ term, denoted $\mathcal{C}_{\lambda_{\Delta_1}}$ and $\mathcal{C}_{\lambda_{\Delta_2}}$, respectively:
\begin{equation}
	\mathcal{C}_{\lambda_{\Delta_1}}\rightarrow \frac{3}{7}\mathcal{C}^2 + \frac{1}{14}\mathcal{C}^3\, ,\, \mathcal{C}_{\lambda_{\Delta_2}} \rightarrow \frac{1}{2}\mathcal{C}^2\, .
\end{equation}
Similar relations are obtained for the other CGCs and then inserted in the model file via the keyword {\it CGCs}. The complete scalar potential entry of the model file then reads
\begin{Verbatim}[numbers=left,xleftmargin=20pt,formatcom=\color{cyan}]
QuarticTerms: {
 '\lambda_1': {Fields : [H*,H,H*,H], Norm : 1, CGCs: {SU2L: [1]}},
 '\lambda_2': {Fields: [T*,T,H*,H], Norm: 1, CGCs: {SU2L: [1]}},
 '\lambda_3': {Fields : [[H*,T,T*,H],[H*,T,T*,H]] Norm : [1/2,1/sqrt(2)], 
	 CGCs: {SU2L: [1,2]}},
 '\lambda_{\Delta_1}': {Fields : [[T*,T,T*,T],[T*,T,T*,T]] Norm : [3/7, 1/14],
	 CGCs: {SU2L: [2,3]}},
 '\lambda_{\Delta_2}'': {Fields : [T*,T,T*,T], Norm : [1/2],
	 CGCs: {SU2L: [2]}}
}
\end{Verbatim}
Note that line 6 of the above snippet exemplifies how to enter terms that are linear combinations of \pyrate CGCs. The complete set of RGEs is given in \ref{app:toymodel1}.

\section{Additional new features}
\label{sec:additional}
In addition to the above described new capabilities, \pyratetwo benefits from all the modifications made to the code since the first release. 
Furthermore, we implemented the anomalous dimensions at two-loop for the scalar and fermion fields
given in \cite{Luo:2002ti}.
%$\beta$-functions for the scalar and fermion fields anomalous dimensions at two-loop from~\cite{Luo:2002ti}. 
Finally, the numerical output has been optimized and it is now possible to solve the RGEs directly in \CC leading to a significant improvement in speed.

\subsection{Scalar and Fermion Fields anomalous dimension at two-loop}
The two-loop RGEs for the scalar and fermion field anomalous dimensions are expressed in terms of the quadratic Casimir operator $C_2$ and the Dynkin 
index $S_2$ of the gauge group which can be related to the generators $t^A$ for fermions and $\Theta^A$ for scalars
\begin{eqnarray}
&C_2^{ab}(S) = \Theta^A_{ac} \Theta^A_{cb}\,,\hspace{1cm} S_2(S)\delta_{AB}= \mbox{Tr}(\Theta^A\Theta^B) \, , &\\
&C_2^{ab}(F) = t^A_{ac} t^A_{cb}\,,\hspace{1cm} S_2(F)\delta_{AB}= \mbox{Tr}(t^A t^B) \, . &
\end{eqnarray}
We also use the Dynkin index	summed over all the states present in the model, $\tilde{S}_{2,k}$ as defined in~\cite{Lyonnet:2013dna}. With these definitions, the anomalous dimensions for a single gauge group $\mathcal{G}_{k}(g)$ read at two-loop~\cite{Luo:2002ti} 
\begin{equation}
\gamma_{ab}^{s} = \gamma_{ab}^{s,I} + \gamma_{ab}^{s,II}\, ,\ \gamma_{ij}^{F} = \gamma_{ij}^{F,I} + \gamma_{ij}^{F,II}\
\end{equation}
with
\begin{eqnarray}
	\gamma_{ab}^{s,I} &=& 2 \kappa Y^{ab}_2(S) - g^{2}(3-\xi) C_{2}^{ab}(S)\, ,\nonumber\\
	\gamma_{ab}^{s,II} &=& -g^{4}C_{2}^{ab}(S)\left[ \left( \frac{35}{3}-2\xi -\frac{1}{4}\xi^{2} \right)C_{2}(\mathcal{G}_{k}) -\frac{10}{3}\kappa \tilde{S}_{2,k}(F) -\frac{11}{12}\tilde{S}_{2,k}(S)\right] +\frac{1}{2}\Lambda_{ab}^{2}(S)\nonumber\\
	&\phantom{.}&+\frac{3}{2}g^{4}C^{ac}_{2}(S) C_{2}^{cb}(S) - 3\kappa H^{2}_{ab}(S)-2\kappa \bar{H}^{2}_{ab}(S) + 10 \kappa g^{2}Y^{2F}_{ab}(S)\, .
\end{eqnarray}
and
\begin{eqnarray}
	\gamma_{ij}^{F,I} &=& \frac{1}{2}Y^{a}_{il}Y^{\dagger a}_{lj} + g^{2}C_{2}^{ij}(F)\xi\, ,\nonumber\\
	\gamma_{ij}^{F,II} &=& -\frac{1}{8}Y^{a}_{il}Y^{\dagger b}_{lm}Y^{b}_{mn}Y^{\dagger a}_{nj} - \frac{3}{2}\kappa Y^{a}_{il}Y^{\dagger b}_{lj}Y^{ab}_2(S)\nonumber\\
	&\phantom{.}& + g^{2}\left[ \frac{9}{2}C_{2}^{ab}(S) Y^{a}_{il}Y^{\dagger b}_{lj} - \frac{7}{4}C_{2}(F)^{il} Y^{a}_{lm}Y^{\dagger a}_{mj} - \frac{1}{4}Y^{a}_{il}C_{2}(F)^{lm}Y^{\dagger b}_{mj} \right]\nonumber\\
	&\phantom{,}& +g^{4}C_{2}(F)^{ij}\left[ \left( \frac{25}{4} + 2\xi +\frac{1}{4}\xi^{2} \right)C_{2}(\mathcal{G}) - 2\kappa \tilde{S}_{2,k}(F) -\frac{1}{4}\tilde{S}_{2,k}(S) \right] -\frac{3}{2}g^{4}C_{2}(F)^{il}C_{2}(F)^{lj}\, ,
\end{eqnarray}
in which $\xi$ is the gauge parameter defined such that $\xi=0$ is the Feynam gauge. 
As usual, $\kappa=\frac{1}{2}(1)$ for	two(four)-component fermions. The definitions of $\Lambda^{2}_{ab}(S),\ H^{2}_{ab}(S),\ \bar{H}^{2}_{ab}(S)$ and $Y^{2F}_{ab}(S)$ are not relevant for our discussion and can be found in~\cite{Luo:2002ti}.

Because the number of possible $a,\ b$ ($i,\ j$) combinations grows rapidly with the number of scalars (fermions) we implemented new key words, \verb|ScalarAnomalous| (\verb|FermionAnomalous|), for the user to specify which combination to calculate. For instance, in the SM, writing $H=(1/\sqrt{2}) (\pi + i \sigma)$ one has 16 combinations:
\begin{equation}
	\gamma(\pi^i,\sigma^j),\ \gamma(\pi^i,\pi^j),\ \gamma(\sigma^i,\pi^j),\ \gamma(\sigma^i,\sigma^j)\, \,
\end{equation}
in which $i,j\in\{1,2\}$ are $\SU{2}_L$ indices. One would therefore enter any or several of these terms directly in the model file e.g., for $\gamma(\pi^1,\sigma^1$)

\begin{Verbatim}[numbers=left,xleftmargin=20pt,formatcom=\color{cyan}]
ScalarAnomalous: {
 '\gamma(\pi,\sigma,1,1)': {Fields : [Pi,Sigma], Norm : 1, 
	 Indices: {SU2L: [1, 1]}},
}
\end{Verbatim}
Several comments are in order here:\begin{inparaenum}[(i)]\item the indices always start from 1, 0 is reserved for a field not charged under the given gauge group; \item \verb|Fields| must contain only real scalars properly defined in the \verb|RealScalars| section or via the \verb|RealFields| property of a \verb|CplxScalar|, see~\cite{Lyonnet:2013dna} for more info; \item fermion anomalous dimensions are specified in the same way via the corresponding \verb|FermionAnomalous| key word.\end{inparaenum}

The whole matrix can be calculated for the scalar by simply omitting the \verb|ScalarAnomalous| entry from the model file. Note that the gauge parameter $\xi$ is kept throughout the calculation such that one can easily set it to some value afterwards.
The two switches controlling the calculation of the anomalous dimensions are \verb|--ScalarAnomalous/-sa| for the scalars and \verb|--FermionAnomalous/-fa| for the fermion fields.

\subsection{{\it Only}, {\it Skip} and GUT normalization}

Three new switches are also making their appearance in \pyratetwo:\begin{inparaenum}[(i)]\item \verb|--Skip/sk|, \item \verb|--Only/onl| and \item \verb|--SetGutNorm/-gutn|\end{inparaenum}. 
As mentioned in \ref{sec:first_steps}, the user now has the possibility to neglect some of the terms 
in the calculation of the RGEs using (i). This is particularly useful when developing a new model or when one wants to neglect tiny terms at the generation level and speed up the calculation of the RGEs. The list of terms that can be neglected this way is based on the grouping usually done in the literature, e.g. \cite{Luo:2002ti}. For instance, the one-loop $\beta$-function for the quartic terms read
$$\beta_{abcd}^{I} = \Lambda^{2}_{abcd} - 8 \kappa H_{abcd} + 2\kappa\Lambda^{Y}_{abcd} -3 g^{2}\Lambda^S_{abcd} + 3 g^4 A_{abcd}\, .$$
Neglecting the terms proportional to the gauge couplings $\Lambda^S_{abcd},\ A_{abcd}$ is achieved by passing the flag
\begin{Verbatim}[numbers=left,xleftmargin=20pt,formatcom=\color{cyan}]
-sk ['CLSabcd','CAabcd']
\end{Verbatim}
Note the capital {\it C} in front of the name of the invariant. Note also that all the possible terms are being printed on the screen when running the calculation with the verbose mode $\verb|-v|$.

The \verb|--Only/-onl| option has a slightly different scope. 
It allows the user to calculate the RGEs for only some of the terms defined in the model file. 
The rest of the Lagrangian is still taken into account for the calculation. The syntax is very similar and one just passes the name of the couplings to include, e.g.
\begin{Verbatim}[numbers=left,xleftmargin=20pt,formatcom=\color{cyan}]
-onl ['\lambda_1','\lambda_2']
\end{Verbatim}

Finally, we introduced the possibility to replace the $\mathrm{U}(1)$ coupling ($g_1$) by the coupling $g^{\prime}$ which has the standard $\mathrm{SU}(5)$ normalization, i.e., $g_1= \sqrt{3/5}g'$. All the RGEs are then expressed in terms of $g'$ instead of $g_1$. This is enabled by passing \verb|--SetGutNorm/-gutn|.
\subsection{Efficient numerical solving of the RGEs}

\pyrate has also been extended by adding a numerical \CC output. 
The set of coupled differential equations is now exported to \CC in addition to the already existing \mathematica and \python outputs. 
The \CC code can then be compiled via the provided {\it Makefile} into a shared library and solved in \python using the routines which have already been provided for this purpose.
This leads to a drastic improvement in speed for solving the RGEs. 
\ref{tab:speedcpp} reports the different times of execution at one- and two-loop. 
These numbers are obtained for the SM+complex triplet model detailed above and solved for 393 points in energy scale between $M_Z$ and $10^{19}$ GeV. 
As one can see, the \CC routine is a factor of almost 9 faster than the corresponding \code{NumPy} one. 
At two-loop, the difference is much bigger reaching a factor of 38. 
However, the time of execution grows linearly for both routines with the number of points in $\log{(t)}$ required.

Note that as in \pyrate, an example file called {\it SolveRGEsCpp.py} is produced with each run and illustrates how to solve the system of $\beta$-functions using the \CC routines. The \CC routines rely on the Armadillo~\cite{armadillo} library for the matrix computation that must be installed independently of \pyrate in order to use this functionality.

\begin{table}
	\centering
	\begin{tabular}{c|c|c}
		times (ms) & \code{NumPy}  & \CC \\\hline\hline
			one-loop & 431  & 48.8 \\ 
			two-loop & 2230	& 58.1\\
	\end{tabular}
	\caption{Time to solve the coupled system of $\beta$-functions at 393 different points in $\log{(t)}$ for the SM + complex triplet at one- and two-loop using the two different numerical routines.}
	\label{tab:speedcpp}
\end{table}

\section{Conclusion}
\label{sec:conclusions}

Renormalization group equations are a key ingredient to extrapolate theories to different energy scales.
Three years ago, we released a \python code, \pyrate, which automatically generates the full two-loop 
RGEs for all the dimensionless and dimensionful parameters of a general gauge theory.
In this article, we have presented the new features implemented in \pyratetwo.
Most importantly, \pyratetwo now supports the kinetic mixing when several \U{1} gauge groups are involved. 
In addition, a new \python module dubbed PyLie has been introduced that 
deals with all the group theoretical aspects required for the calculation of the RGEs and provides
several functions useful for model building. In particular, any irreducible representation of the 
$\mathrm{SU}(n)$, $\mathrm{SO}(2n)$ and $\mathrm{SO(2n+1)}$ groups is now supported.
Furthermore, it is now possible to handle combinations of fields which can be contracted into gauge singlets
in multiple ways.
Finally, the two-loop RGEs for the anomalous dimensions of the scalar and fermion fields have been 
implemented as well. It is now possible to export the coupled system of beta functions into a 
numerical \CC function, leading to a consequent speed up in solving them.

\section*{Acknowledgments}
We are grateful to Florian Staub, Renato Fonseca, Kristjan Kannike, Helena Kole\v{s}ov\'{a} and Fred Olness for many useful discussions. 
F.L. would like to thank Florian Staub for helping validating the implementation of the kinetic mixing. F.L. is also greatful to Luigi Delle Rose for providing insights on the implementation of the kinetic mixing and for the help in resolving early discrepencies in the $\U{1}_{B-L}$ model.

\appendix
\clearpage
\newpage
\labelformat{section}{#1}
\newpage
\section{SM-$\mathrm{U(1)}_{B-L}$}
%%% TODO Yukawa and scalar mass at two loop agreed with Luigi and so I copied them here The  quartic couplings need to be updated
\label{app:twobl}
\subsection{Yukawa couplings beta function}
\begin{eqnarray}
	\beta^{(2)}_{Y_t} &=&\frac{1}{432} Y_t \left(4 \tilde{g} g' \left(270 \text{Tr}\left[Y_{\nu }Y_{\nu }{}^{\dagger }\right]+2008 g^2+2660 \left(g'\right)^2+243 g_2^2-240 g_3^2\right.\right.\nonumber\\
	&\phantom{.}&\left.+450 Y_t^2+225 Y_t^2\right)+\tilde{g}^2 \left(270 \text{Tr}\left[Y_{\nu }Y_{\nu }{}^{\dagger }\right]+4748 g^2\right.\nonumber\\
	&\phantom{.}&\left.+13020 \left(g'\right)^2-324 g_2^2+912 g_3^2+1530 Y_t^2+2007 Y_t^2\right)+8032 \tilde{g}^3 g'+2374 \tilde{g}^4\nonumber\\
	&\phantom{.}&+270 g^2 \text{Tr}\left[Y_{\nu }Y_{\nu }{}^{\dagger }\right]+2160 \left(g'\right)^2 \text{Tr}\left[Y_{\nu }Y_{\nu }{}^{\dagger }\right]-81 g_2^2 \left(-10 \text{Tr}\left[Y_{\nu }Y_{\nu }{}^{\dagger }\right]\right.\nonumber\\
	&\phantom{.}&\left.+4 g^2-4 \left(g'\right)^2-48 g_3^2-30 Y_t^2-45 Y_t^2\right)-1296 Y_N^2 \text{Tr}\left[Y_{\nu }{}^{\dagger }Y_{\nu }\right]\nonumber\\
	&\phantom{.}&-972 \text{Tr}\left[Y_{\nu }Y_{\nu }{}^{\dagger }Y_{\nu }Y_{\nu }{}^{\dagger }\right]-972 Y_t^2 \text{Tr}\left[Y_{\nu }Y_{\nu }{}^{\dagger }\right]+2374 g^4\nonumber\\
	&\phantom{.}&+1530 g^2 Y_t^2+2007 g^2 Y_t^2+912 g_3^2 g^2+720 \left(g'\right)^2 Y_t^2+576 Y_t^2 \left(g'\right)^2+3248 \left(g'\right)^4\nonumber\\
	&\phantom{.}&-384 g_3^2 \left(g'\right)^2+3276 g^2 \left(g'\right)^2+8640 g_3^2 Y_t^2+6912 g_3^2 Y_t^2-2484 g_2^4-46656 g_3^4\nonumber\\
	&\phantom{>}&\left.+2592 \lambda _1^2+216 \lambda _3^2-2916 Y_t^2 Y_t^2-2916 Y_t^4+648 Y_t^4-5184 \lambda _1 Y_t^2\right)\\
	\beta^{(2)}_{Y_N} &=& -\frac{1}{24} Y_N \left(2 \left(128 \tilde{g} \left(g'\right)^3+70 \tilde{g}^2 \left(g'\right)^2+12 Y_N^2 \left(Y_{\nu }Y_{\nu }^\dagger\right)-48 \left(Y_{\nu }^tY_{\nu }^* Y_{\nu }{}^{\dagger }Y_{\nu }\right)+3 \left(Y_{\nu }^tY_{\nu }^*Y_{\nu }^tY_{\nu }^*\right)\right.\right.\nonumber\\
	&\phantom{>}&+3 \left(\left(Y_{\nu }\right){}^{\dagger }Y_{\nu }\left(Y_{\nu }\right){}^{\dagger }Y_{\nu }\right)+72 Y_N^2 \text{Tr}\left(\left(Y_{\nu }\right){}^{\dagger },Y_{\nu }\right)-2472 \left(g'\right)^2 Y_N^2\nonumber\\
	&\phantom{>}&\left.+1524 \left(g'\right)^4-48 \lambda _2^2-12 \lambda _3^2+384 \lambda _2 Y_N^2+528 Y_N^4\right)+3 \left(Y_{\nu }^t\left(Y_{\nu }^*\right)\right) \left(52 \tilde{g} g'-17 \tilde{g}^2\right.\nonumber\\
	&\phantom{.}&\left.+12 \text{Tr}\left(Y_{\nu }Y_{\nu }{}^{\dagger }\right)-17 g^2+128 \left(g'\right)^2-51 g_2^2+32 \lambda _3+36 Y_t^2\right)\nonumber\\
	&\phantom{.}&+3 \left(Y_{\nu }{}^{\dagger }Y_{\nu }\right) \left(52 \tilde{g} g'-17 \tilde{g}^2+12 \text{Tr}\left(Y_{\nu }Y_{\nu }{}^{\dagger }\right)-17 g^2+128 \left(g'\right)^2-51 g_2^2\right.\nonumber\\
	&\phantom{.}&\left.\left.+32 \lambda _3+8 Y_N^2+36 Y_t^2\right)\right)\\
	\beta^{(2)}_{Y_\nu} &=& \frac{1}{48} 3 \left(Y_{\nu }Y_{\nu }{}^{\dagger }Y_{\nu }\right) \left(156 \tilde{g} g'+93 \tilde{g}^2-36 \text{Tr}\left(Y_{\nu }Y_{\nu }{}^{\dagger }\right)+93 g^2+192 \left(g'\right)^2+135 g_2^2\right.\nonumber\\
	&\phantom{.}&\left.-192 \lambda _1-8 Y_N^2-108 Y_t^2\right)+2 \left(Y_{\nu } \left(2 \tilde{g} g' \left(30 \text{Tr}\left(Y_{\nu }Y_{\nu }{}^{\dagger }\right)+252 g^2+1012 \left(g'\right)^2\right.\right.\right.\nonumber\\
	&\phantom{.}&\left.+27 g_2^2-144 Y_N^2+50 Y_t^2\right)+\tilde{g}^2 \left(15 \text{Tr}\left(Y_{\nu }Y_{\nu }{}^{\dagger }\right)+70 g^2+1598 \left(g'\right)^2-54 g_2^2+85 Y_t^2\right)\nonumber\\
	&\phantom{.}&+504 \tilde{g}^3 g'+35 \tilde{g}^4+15 g^2 \text{Tr}\left(Y_{\nu }Y_{\nu }{}^{\dagger }\right)+120 \left(g'\right)^2 \text{Tr}\left(Y_{\nu }Y_{\nu }{}^{\dagger }\right)+9 g_2^2 \left(5 \text{Tr}\left(Y_{\nu }Y_{\nu }{}^{\dagger }\right)\right.\nonumber\\
	&\phantom{>}&\left.-6 g^2+18 \left(g'\right)^2+15 Y_t^2\right)-72 Y_N^2 \text{Tr}\left(Y_{\nu }{}^{\dagger }Y_{\nu }\right)-54 \text{Tr}\left(Y_{\nu }Y_{\nu }{}^{\dagger }Y_{\nu }Y_{\nu }{}^{\dagger }\right)+35 g^4+85 g^2 Y_t^2\nonumber\\
	&\phantom{.}&+960 \left(g'\right)^2 Y_N^2+40 \left(g'\right)^2 Y_t^2+1560 \left(g'\right)^4+374 g^2 \left(g'\right)^2+480 g_3^2 Y_t^2-138 g_2^4+144 \lambda_1^2\nonumber\\
	&\phantom{.}&\left.\left.+12 \lambda _3^2-192 \lambda_3 Y_N^2-480 Y_N^4-162 Y_t^4\right)+168 Y_N^2 \left(Y_{\nu }Y_{\nu }^tY_{\nu }^*\right)+36 \left(Y_{\nu }Y_{\nu }{}^{\dagger }Y_{\nu }Y_{\nu }{}^{\dagger }Y_{\nu }\right)\right)
\end{eqnarray}
\subsection{Scalar mass beta function}
\begin{eqnarray}
	\beta^{(2)}_{\mu} &=&\frac{1}{48} \tilde{g}^2 \left(60 \mu  \text{Tr}\left[Y_{\nu }Y_{\nu }{}^{\dagger }\right]+1105 g^2 \mu +960 \left(g'\right)^2 \mu _{\chi }+816 \mu  \left(g'\right)^2+63 g_2^2 \mu +1152 \lambda _1 \mu +340 \mu  Y_t^2\right)\nonumber\\
	&\phantom{.}&+\frac{5}{3} \mu  \tilde{g} g' \left(3 \text{Tr}\left[Y_{\nu }Y_{\nu }{}^{\dagger }\right]+8 g^2+5 Y_t^2\right)+\frac{3}{16} g_2^2 \mu  \left(20 \text{Tr}\left[Y_{\nu }Y_{\nu }{}^{\dagger }\right]+7 g^2+384 \lambda _1+60 Y_t^2\right)\nonumber\\
	&\phantom{.}&+\frac{5}{4} g^2 \mu  \text{Tr}\left[Y_{\nu }Y_{\nu }{}^{\dagger }\right]+10 \mu  \left(g'\right)^2 \text{Tr}\left[Y_{\nu }Y_{\nu }{}^{\dagger }\right]-6 \mu  Y_N^2 \text{Tr}\left[Y_{\nu }{}^{\dagger }Y_{\nu }\right]-24 \lambda _1 \mu  \text{Tr}\left[Y_{\nu }Y_{\nu }{}^{\dagger }\right]\nonumber\\
	&\phantom{>}&+\frac{40}{3} \mu  \tilde{g}^3 g'-\frac{9}{2} \mu  \text{Tr}\left[Y_{\nu }Y_{\nu }{}^{\dagger }Y_{\nu }Y_{\nu }{}^{\dagger }\right]+\frac{1105 g^4 \mu }{96}+24 g^2 \lambda _1 \mu +\frac{85}{12} g^2 \mu  Y_t^2+64 \lambda _3 \left(g'\right)^2 \mu _{\chi }\nonumber\\
	&\phantom{.}&+\frac{10}{3} \mu  \left(g'\right)^2 Y_t^2-\frac{145 g_2^4 \mu }{16}+40 g_3^2 \mu  Y_t^2-4 \lambda _3^2 \mu _{\chi }-60 \lambda _1^2 \mu -\lambda _3^2 \mu -24 \lambda _3 \mu _{\chi } Y_N^2-72 \lambda _1 \mu  Y_t^2\nonumber\\
	&\phantom{.}&-\frac{27 \mu  Y_t^4}{2} +\frac{1105 \mu  \tilde{g}^4}{96}\\
\beta^{(2)}_{\mu_{\chi}} &=&88 \lambda _3 \mu \left(\tilde{g}^2-\text{Tr}\left[Y_{\nu }Y_{\nu }{}^{\dagger }\right]+g^2+3 g_2^2-3 Y_t^2\right)+\frac{640}{3} \tilde{g} \left(g'\right)^3 \mu _{\chi }+\frac{422}{3} \tilde{g}^2 \left(g'\right)^2 \mu _{\chi }\nonumber\\
&\phantom{.}&+40 \mu  \tilde{g}^2 \left(g'\right)^2-12 \mu _{\chi } Y_N^2 \left(\text{Tr}\left[Y_{\nu }{}^{\dagger }Y_{\nu }\right]-5 \left(g'\right)^2+8 \lambda _2\right)-24 \mu _{\chi } \text{Tr}\left[Y_NY_N{}^{\dagger }Y_NY_N{}^{\dagger }\right]\nonumber\\
&\phantom{>}&+256 \lambda _2 \left(g'\right)^2 \mu _{\chi }+648 \left(g'\right)^4 \mu _{\chi }-40 \lambda _2^2 \mu _{\chi }-2 \lambda _3^2 \left(\mu _{\chi }+4 \mu \right)
\end{eqnarray}
\subsection{Quartic couplings beta function}
We present here the complete list of the two-loops terms for the beta functions of the quartic terms for the SM-$\mathrm{U}(1)_{B-L}$. Note that all the replacement rules of~\cite{Fonseca:2013jra} have been implemented.
\begin{eqnarray}
	\beta^{(2)}_{\lambda_1} &=&-\frac{379 g^6}{48}-\frac{19}{4} Y_t^2 g^4+\frac{629 \lambda _1 g^4}{24}-\frac{1}{4} \text{Tr}\left[Y_{\nu }Y_{\nu }{}^{\dagger }\right] g^4-\frac{8}{3} Y_t^4 g^2+36 \lambda _1^2 g^2+\frac{85}{6} Y_t^2 \nonumber\\
	&\phantom{.}&\lambda _1 g^2+\frac{5}{2} \lambda _1 \text{Tr}\left[Y_{\nu }Y_{\nu }{}^{\dagger }\right] g^2+\frac{305 g_2^6}{16}+30 Y_t^6-32 g_3^2 Y_t^4\nonumber\\
	&\phantom{.}&-312 \lambda _1^3-4 \lambda _3^3-144 Y_t^2 \lambda _1^2-12 Y_N^2 \lambda _3^2-10 \lambda _1 \lambda _3^2-\frac{8}{3} Y_t^4 \left(g'\right)^2\nonumber\\
	&\phantom{.}&+32 \lambda _3^2 \left(g'\right)^2+\frac{20}{3} Y_t^2 \lambda _1 \left(g'\right)^2+20 \lambda _1 \text{Tr}\left[Y_{\nu }Y_{\nu }{}^{\dagger }\right] \left(g'\right)^2\nonumber\\
	&\phantom{.}&-8 \text{Tr}\left[Y_{\nu }Y_{\nu }{}^{\dagger }Y_{\nu }Y_{\nu }{}^{\dagger }\right] \left(g'\right)^2-3 Y_t^4 \lambda _1+80 g_3^2 Y_t^2 \lambda _1\nonumber\\
	&\phantom{.}&-48 \lambda _1^2 \text{Tr}\left[Y_{\nu }Y_{\nu }{}^{\dagger }\right]-\frac{1}{48} g_2^4 \left(289 g^2+108 Y_t^2+438 \lambda _1+36 \text{Tr}\left[Y_{\nu }Y_{\nu }{}^{\dagger }\right]\right)\nonumber\\
	&\phantom{.}&-\frac{1}{48} g_2^2 \left(559 g^4+24 \text{Tr}\left[Y_{\nu }Y_{\nu }{}^{\dagger }\right] g^2\right.\nonumber\\
	&\phantom{.}&\left.-5184 \lambda _1^2-72 Y_t^2 \left(7 g^2+15 \lambda _1\right)-36 \lambda _1 \left(13 g^2+10 \text{Tr}\left[Y_{\nu }Y_{\nu }{}^{\dagger }\right]\right)\right)-12 Y_N^2 \lambda _1 \text{Tr}\left[Y_{\nu }{}^{\dagger }Y_{\nu }\right]\nonumber\\
	&\phantom{.}&-\lambda _1 \text{Tr}\left[Y_{\nu }Y_{\nu }{}^{\dagger }Y_{\nu }Y_{\nu }{}^{\dagger }\right]+8 Y_N^2 \text{Tr}\left[Y_{\nu }^tY_{\nu }^*Y_{\nu }{}^{\dagger }Y_{\nu }\right]+8 Y_N^2 \text{Tr}\left[Y_{\nu }{}^{\dagger }Y_{\nu }Y_{\nu }{}^{\dagger }Y_{\nu }\right]\nonumber\\
	&\phantom{>}&+10 \text{Tr}\left[Y_{\nu }Y_{\nu }{}^{\dagger }Y_{\nu }Y_{\nu }{}^{\dagger }Y_{\nu }Y_{\nu }{}^{\dagger }\right]-\frac{32}{3} \tilde{g}^5 g'\nonumber\\
	&\phantom{>}&-\frac{2}{3} \tilde{g}^3 \left(32 g^2+16 g_2^2+15 Y_t^2-40 \lambda _1+9 \text{Tr}\left[Y_{\nu }Y_{\nu }{}^{\dagger }\right]\right) g'-\frac{2}{3} \tilde{g} \left(16 g^4-40 \lambda _1 g^2+9 \text{Tr}\left[Y_{\nu }Y_{\nu }{}^{\dagger }\right] g^2\right.\nonumber\\
	&\phantom{>}&+10 Y_t^4+5 Y_t^2 \left(3 g^2-5 \lambda _1\right)-15 \lambda _1 \text{Tr}\left[Y_{\nu }Y_{\nu }{}^{\dagger }\right]+g_2^2 \left(16 g^2-9 Y_t^2+9 \text{Tr}\left[Y_{\nu }Y_{\nu }{}^{\dagger }\right]\right)\nonumber\\
	&\phantom{.}&\left.+6 \text{Tr}\left[Y_{\nu }Y_{\nu }{}^{\dagger }Y_{\nu }Y_{\nu }{}^{\dagger }\right]\right) g'-\frac{1}{48} \tilde{g}^4 \left(1137 g^2+559 g_2^2+228 Y_t^2+624 \left(g'\right)^2-1258 \lambda _1\right.\nonumber\\
	&\phantom{.}&\left.+12 \text{Tr}\left[Y_{\nu }Y_{\nu }{}^{\dagger }\right]\right)-\frac{1}{48} \tilde{g}^2 \left(1137 g^4+624 \left(g'\right)^2 g^2-2516 \lambda _1 g^2+24 \text{Tr}\left[Y_{\nu }Y_{\nu }{}^{\dagger }\right] g^2\right.\nonumber\\
	&\phantom{>}&+289 g_2^4+128 Y_t^4-1728 \lambda _1^2-1632 \lambda _1 \left(g'\right)^2-960 \lambda _3 \left(g'\right)^2+576 \text{Tr}\left[Y_{\nu }Y_{\nu }{}^{\dagger }\right] \left(g'\right)^2\nonumber\\
	&\phantom{>}&-120 \lambda _1 \text{Tr}\left[Y_{\nu }Y_{\nu }{}^{\dagger }\right]+8 Y_t^2 \left(57 g^2+24 \left(g'\right)^2-85 \lambda _1\right)\nonumber\\
	&\phantom{.}&\left.+2 g_2^2 \left(559 g^2-252 Y_t^2+312 \left(g'\right)^2-234 \lambda _1+12 \text{Tr}\left[Y_{\nu }Y_{\nu }{}^{\dagger }\right]\right)\right)-\frac{379 \tilde{g}^6}{48}
\end{eqnarray}
\begin{eqnarray}
	\beta^{(2)}_{\lambda_2} &=&-\frac{1}{3} 713 \left(g'\right)^2 \tilde{g}^4-\frac{1024}{3} \left(g'\right)^3 \tilde{g}^3-656 \left(g'\right)^4 \tilde{g}^2-\frac{713}{3} g^2 \left(g'\right)^2 \tilde{g}^2-45 g_2^2 \left(g'\right)^2 \tilde{g}^2\nonumber\\
	&\phantom{>}&-76 Y_t^2 \left(g'\right)^2 \tilde{g}^2+120 \lambda _1 \left(g'\right)^2 \tilde{g}^2+80 \lambda _2 \left(g'\right)^2 \tilde{g}^2\nonumber\\
	&\phantom{>}&-4 \text{Tr}\left[Y_{\nu }Y_{\nu }{}^{\dagger }\right] \left(g'\right)^2 \tilde{g}^2+\text{Tr}\left[Y_NY_{\nu }{}^{\dagger }Y_{\nu }Y_N{}^{\dagger }\right] \tilde{g}^2-\frac{512}{3} g^2 \left(g'\right)^3 \tilde{g}-160 Y_t^2 \left(g'\right)^3 \tilde{g}\nonumber\\
	&\phantom{.}&-96 \text{Tr}\left[Y_{\nu }Y_{\nu }{}^{\dagger }\right] \left(g'\right)^3 \tilde{g}-64 Y_t^2 \left(g'\right)^4-192 \text{Tr}\left[Y_{\nu }Y_{\nu }{}^{\dagger }\right] \left(g'\right)^4-11 \lambda _3^3+g^2 \text{Tr}\left[Y_NY_{\nu }{}^{\dagger }Y_{\nu }Y_N{}^{\dagger }\right]\nonumber\\
	&\phantom{.}&+3 g_2^2 \text{Tr}\left[Y_NY_{\nu }{}^{\dagger }Y_{\nu }Y_N{}^{\dagger }\right]+32 \text{Tr}\left[Y_N^*Y_N^tY_N^*Y_N^tY_{\nu }{}^{\dagger }Y_{\nu }\right]+32 \text{Tr}\left[Y_N^*Y_N^tY_{\nu }{}^{\dagger }Y_{\nu }Y_N^*Y_N^t\right]\nonumber\\
	&\phantom{.}&+16 \text{Tr}\left[Y_N^*Y_N^tY_{\nu }{}^{\dagger }Y_{\nu }Y_{\nu }{}^{\dagger }Y_{\nu }\right]+64 \text{Tr}\left[Y_NY_N{}^{\dagger }Y_NY_{\nu }{}^{\dagger }Y_{\nu }Y_N{}^{\dagger }\right]+32 \text{Tr}\left[Y_NY_{\nu }{}^{\dagger }Y_{\nu }Y_N{}^{\dagger }Y_NY_N{}^{\dagger }\right] \nonumber\\
	&\phantom{>}&+\lambda _3^2 \left(g^2+\tilde{g}^2+3 g_2^2-24 Y_N^2-12 Y_t^2+16 \left(g'\right)^2-72 \lambda _1-48 \lambda _2-4 \text{Tr}\left[Y_{\nu }Y_{\nu }{}^{\dagger }\right]\right)\nonumber\\
	&\phantom{.}&+\frac{1}{48} \lambda _3 \left(557 g^4+340 Y_t^2 g^2+1152 \lambda _1 g^2+60 \text{Tr}\left[Y_{\nu }Y_{\nu }{}^{\dagger }\right] g^2+557 \tilde{g}^4-435 g_2^4-648 Y_t^4+32256 \left(g'\right)^4\right.\nonumber\\
		&\phantom{.}&+1920 g_3^2 Y_t^2-2880 \lambda _1^2-1920 \lambda _2^2+2880 Y_N^2 \left(g'\right)^2+160 Y_t^2 \left(g'\right)^2+12288 \lambda _2 \left(g'\right)^2+480 \text{Tr}\left[Y_{\nu }Y_{\nu }{}^{\dagger }\right] \left(g'\right)^2\nonumber\\
		&\phantom{>}&-3456 Y_t^2 \lambda _1-4608 Y_N^2 \lambda _2-1152 \lambda _1 \text{Tr}\left[Y_{\nu }Y_{\nu }{}^{\dagger }\right]+18 g_2^2 \left(5 g^2+30 Y_t^2+192 \lambda _1+10 \text{Tr}\left[Y_{\nu }Y_{\nu }{}^{\dagger }\right]\right)\nonumber\\
		&\phantom{.}&+288 Y_N^2 \text{Tr}\left[Y_{\nu }{}^{\dagger }Y_{\nu }\right]-1152 \text{Tr}\left[Y_NY_N{}^{\dagger }Y_NY_N{}^{\dagger }\right]+192 \text{Tr}\left[Y_NY_{\nu }{}^{\dagger }Y_{\nu }Y_N{}^{\dagger }\right]+192 \text{Tr}\left[Y_N^tY_{\nu }{}^{\dagger }Y_{\nu }Y_N^*\right]\nonumber\\
		&\phantom{>}&-216 \text{Tr}\left[Y_{\nu }Y_{\nu }{}^{\dagger }Y_{\nu }Y_{\nu }{}^{\dagger }\right]+640 \tilde{g}^3 g'+80 \tilde{g} g' \left(8 g^2+5 Y_t^2+128 \left(g'\right)^2+3 \text{Tr}\left[Y_{\nu }Y_{\nu }{}^{\dagger }\right]\right)\nonumber\\
		&\phantom{>}&\left.+2 \tilde{g}^2 \left(557 g^2+45 g_2^2+170 Y_t^2+3976 \left(g'\right)^2+576 \lambda _1+30 \text{Tr}\left[Y_{\nu }Y_{\nu }{}^{\dagger }\right]\right)\right)\nonumber\\
		&\phantom{>}&-Y_N^2 \left(\text{Tr}\left[Y_{\nu }{}^{\dagger }Y_{\nu }\right] \left(g^2+\tilde{g}^2+3 g_2^2+16 \tilde{g} g'\right)\right.\nonumber\\
		&\phantom{.}&\left.-8 \left(6 \tilde{g}^2 \left(g'\right)^2+4 \text{Tr}\left[Y_{\nu }^tY_{\nu }^*Y_{\nu }{}^{\dagger }Y_{\nu }\right]+5 \text{Tr}\left[Y_{\nu }{}^{\dagger }Y_{\nu }Y_{\nu }{}^{\dagger }Y_{\nu }\right]\right)\right)
\end{eqnarray}

\begin{eqnarray}
	\beta_{\lambda_3}^{(2)} &=&\frac{1}{3} (-4) \left(\lambda _2 \left(-\left(g'\right)^2 \left(320 \tilde{g} g'+211 \tilde{g}^2+1584 \left(g'\right)^2\right)-12 \text{Tr}\left[Y_NY_N{}^{\dagger }Y_NY_N{}^{\dagger }\right]+15 \lambda _3^2\right)-30 \lambda _3 \tilde{g}^2 \left(g'\right)^2\right.\nonumber\\
	&\phantom{.}&+2048 \tilde{g} \left(g'\right)^5+1336 \tilde{g}^2 \left(g'\right)^4-3 \lambda _3^2 \tilde{g}^2+18 Y_N^2 \left(\lambda _2 \left(\text{Tr}\left[Y_{\nu }{}^{\dagger }Y_{\nu }\right]-5 \left(g'\right)^2\right)-32 \left(g'\right)^4\right.\nonumber\\
	&\phantom{.}&\left.+10 \lambda _2^2\right)-48 \left(g'\right)^2 \text{Tr}\left[Y_NY_N{}^{\dagger }Y_NY_N{}^{\dagger }\right]-48 \text{Tr}\left[Y_NY_N{}^{\dagger }Y_NY_{\nu }{}^{\dagger }Y_{\nu }Y_N{}^{\dagger }\right]\nonumber\\
	&\phantom{.}&-192 \text{Tr}\left[Y_NY_N{}^{\dagger }Y_NY_N{}^{\dagger }Y_NY_N{}^{\dagger }\right]+3 \lambda _3^2 \text{Tr}\left[Y_{\nu }Y_{\nu }{}^{\dagger }\right]-3 g^2 \lambda _3^2-336 \lambda _2^2 \left(g'\right)^2+5376 \left(g'\right)^6-9 g_2^2 \lambda _3^2\nonumber\\
	&\phantom{.}&\left.+180 \lambda _2^3+6 \lambda _3^3+9 \lambda _3^2 Y_t^2\right) 
\end{eqnarray}

\section{RGEs for the SM+complex triplet}
\label{app:toymodel1}
Here we list the full set of RGEs for the SM + complex triplet model including the two invariants coming from the contraction of four triplet fields defined by the potential in \ref{eq:potentialToyTII}.

			\subsection{Gauge Couplings}
				\begin{align*}
					\frac{dg}{dt} = \left.\beta_g\right|_I + \left.\beta_g\right|_{II}\
				\end{align*}
					\subsubsection{Evolution of $g_{2}$}
					\begin{align*}
						\left.(4\pi)^2\beta_{g_{2}}\right|_{I} = 
					&- \frac{3}{2} g_{2}^{3}\\
						\left.(4\pi)^4\beta_{g_{2}}\right|_{II} = 
				&+\frac{41}{4} g_{1}^{2} g_{2}^{3} + \frac{147}{4} g_{2}^{5} + 12 g_{2}^{3} g_{3}^{2}\\
					\end{align*}
					\subsubsection{Evolution of $g_{1}$}
					\begin{align*}
						\left.(4\pi)^2\beta_{g_{1}}\right|_{I} = 
					&+\frac{53}{6} g_{1}^{3}\\
						\left.(4\pi)^4\beta_{g_{1}}\right|_{II} = 
				&+\frac{857}{36} g_{1}^{5} + \frac{123}{4} g_{1}^{3} g_{2}^{2} + \frac{44}{3} g_{1}^{3} g_{3}^{2}\\
					\end{align*}
					\subsubsection{Evolution of $g_{3}$}
					\begin{align*}
						\left.(4\pi)^2\beta_{g_{3}}\right|_{I} = 
					&- 7 g_{3}^{3}\\
						\left.(4\pi)^4\beta_{g_{3}}\right|_{II} = 
				&+\frac{11}{6} g_{1}^{2} g_{3}^{3} + \frac{9}{2} g_{2}^{2} g_{3}^{3} - 26 g_{3}^{5}\\
					\end{align*}
	
		\subsection{Quartic Coupling}
		\begin{align*}
			\frac{d\lambda}{dt} = \left.\beta_{\lambda}\right|_I + \left.\beta_{\lambda}\right|_{II}\
		\end{align*}
			\subsubsection{Evolution of $\lambda_{1}$}
				\begin{align*}
					\left.(4\pi)^2\beta_{\lambda_{1}}\right|_{I}=
				&+24 \lambda_{1}^{2} + \frac{5}{4} \lambda_{{3}}^{2} + 3 \lambda_{{3}} \lambda_{{2}} + 3 \lambda_{{2}}^{2} + \frac{9}{8} g_{2}^{4}\\
				&- 3 \lambda_{1} g_{1}^{2} - 9 \lambda_{1} g_{2}^{2} + \frac{3}{8} g_{1}^{4} + \frac{3}{4} g_{1}^{2} g_{2}^{2}\\\end{align*}\begin{align*}
			\left.(4\pi)^4\beta_{\lambda_{1}}\right|_{II}=
				&- 312 \lambda_{1}^{3} + 15 \lambda_{{2}} g_{1}^{4} + 30 \lambda_{{2}} g_{2}^{4} - \frac{511}{48} g_{1}^{6}\\
				&+\frac{821}{24} \lambda_{1} g_{1}^{4} + 10 \lambda_{{3}}^{2} g_{1}^{2} - \frac{691}{48} g_{1}^{4} g_{2}^{2} + \frac{201}{16} g_{2}^{6}\\
				&+17 \lambda_{{3}}^{2} g_{2}^{2} - 10 \lambda_{{3}} g_{1}^{2} g_{2}^{2} + 24 \lambda_{{2}}^{2} g_{1}^{2}\\
				&- 30 \lambda_{1} \lambda_{{3}} \lambda_{{2}} - 30 \lambda_{1} \lambda_{{2}}^{2} + 24 \lambda_{{3}} \lambda_{{2}} g_{1}^{2}\\
				&+108 \lambda_{1}^{2} g_{2}^{2} - \frac{29}{2} \lambda_{1} \lambda_{{3}}^{2} - 12 \lambda_{{2}}^{3}\\
				&- 18 \lambda_{{3}} \lambda_{{2}}^{2} + 15 \lambda_{{3}} g_{2}^{4} - \frac{131}{16} g_{1}^{2} g_{2}^{4}\\
				&- 19 \lambda_{{3}}^{2} \lambda_{{2}} + \frac{15}{2} \lambda_{{3}} g_{1}^{4} + 48 \lambda_{{2}}^{2} g_{2}^{2}\\
				&+36 \lambda_{1}^{2} g_{1}^{2} - \frac{13}{2} \lambda_{{3}}^{3} + 48 \lambda_{{3}} \lambda_{{2}} g_{2}^{2}\\
				&+\frac{39}{4} \lambda_{1} g_{1}^{2} g_{2}^{2} + \frac{75}{8} \lambda_{1} g_{2}^{4}\\
			\end{align*}
			\subsubsection{Evolution of $\lambda_{{3}}$}
				\begin{align*}
					\left.(4\pi)^2\beta_{\lambda_{{3}}}\right|_{I}=
				&+8 \lambda_{{3}} \lambda_{{2}} - \frac{15}{2} \lambda_{{3}} g_{1}^{2} - \frac{33}{2} \lambda_{{3}} g_{2}^{2} - 12 g_{1}^{2} g_{2}^{2}\\
				&+4 \lambda_{1} \lambda_{{3}} + 4 \lambda_{{3}}^{2} + 8 \lambda_{{3}} \lambda_{{\Delta 2}} + 4 \lambda_{{3}} \lambda_{{\Delta 1}}\\\end{align*}\begin{align*}
			\left.(4\pi)^4\beta_{\lambda_{{3}}}\right|_{II}=
				&- \frac{27}{4} \lambda_{{3}}^{3} - 96 \lambda_{{3}} \lambda_{{2}} \lambda_{{\Delta 1}} - 8 \lambda_{{2}} g_{1}^{2} g_{2}^{2}\\
				&+64 \lambda_{{3}} \lambda_{{\Delta 2}} g_{1}^{2} + 46 \lambda_{{3}} \lambda_{{2}} g_{2}^{2} + \frac{3421}{48} \lambda_{{3}} g_{1}^{4}\\
				&- 28 \lambda_{1}^{2} \lambda_{{3}} - \frac{245}{16} \lambda_{{3}} g_{2}^{4} - 80 \lambda_{{\Delta 2}} g_{1}^{2} g_{2}^{2}\\
				&- 40 \lambda_{1} \lambda_{{3}}^{2} - 48 \lambda_{{3}}^{2} \lambda_{{\Delta 1}} - \lambda_{{3}}^{2} g_{1}^{2}\\
				&- 80 \lambda_{1} \lambda_{{3}} \lambda_{{2}} - 40 \lambda_{{\Delta 1}} g_{1}^{2} g_{2}^{2} + \frac{691}{3} g_{1}^{4} g_{2}^{2}\\
				&- 38 \lambda_{{3}} \lambda_{{\Delta 2}}^{2} + 80 \lambda_{{3}} \lambda_{{\Delta 2}} g_{2}^{2} - 29 \lambda_{{3}} \lambda_{{2}}^{2}\\
				&+23 \lambda_{{3}}^{2} g_{2}^{2} - 112 \lambda_{{3}} \lambda_{{\Delta 2}} \lambda_{{2}} + 32 \lambda_{{3}} \lambda_{{\Delta 1}} g_{1}^{2}\\
				&- 40 \lambda_{1} g_{1}^{2} g_{2}^{2} - 29 \lambda_{{3}}^{2} \lambda_{{2}} + 40 \lambda_{{3}} \lambda_{{\Delta 1}} g_{2}^{2}\\
				&+8 \lambda_{1} \lambda_{{3}} g_{1}^{2} - 32 \lambda_{{3}} \lambda_{{\Delta 1}}^{2} + \frac{783}{8} \lambda_{{3}} g_{1}^{2} g_{2}^{2}\\
				&- 56 \lambda_{{3}}^{2} \lambda_{{\Delta 2}} - 88 \lambda_{{3}} \lambda_{{\Delta 2}} \lambda_{{\Delta 1}} + 10 \lambda_{{3}} \lambda_{{2}} g_{1}^{2}\\
				&+161 g_{1}^{2} g_{2}^{4}\\
			\end{align*}
			\subsubsection{Evolution of $\lambda_{{2}}$}
				\begin{align*}
					\left.(4\pi)^2\beta_{\lambda_{{2}}}\right|_{I}=
				&+12 \lambda_{{\Delta 2}} \lambda_{{2}} + 16 \lambda_{{2}} \lambda_{{\Delta 1}} - \frac{15}{2} \lambda_{{2}} g_{1}^{2}\\
				&+4 \lambda_{1} \lambda_{{3}} + 2 \lambda_{{3}} \lambda_{{\Delta 2}} - \frac{33}{2} \lambda_{{2}} g_{2}^{2} + 3 g_{1}^{4} + 6 g_{1}^{2} g_{2}^{2}\\
				&+12 \lambda_{1} \lambda_{{2}} + \lambda_{{3}}^{2} + 6 \lambda_{{3}} \lambda_{{\Delta 1}} + 4 \lambda_{{2}}^{2} + 6 g_{2}^{4}\\\end{align*}\begin{align*}
			\left.(4\pi)^4\beta_{\lambda_{{2}}}\right|_{II}=
				&+60 \lambda_{1} g_{2}^{4} - 72 \lambda_{{\Delta 2}} \lambda_{{2}}^{2} - 96 \lambda_{{2}}^{2} \lambda_{{\Delta 1}}\\
				&- 5 \lambda_{{3}} \lambda_{{2}}^{2} - 12 \lambda_{{3}} \lambda_{{2}} g_{2}^{2} + 40 \lambda_{{\Delta 2}} g_{1}^{2} g_{2}^{2}\\
				&+20 \lambda_{1} g_{1}^{2} g_{2}^{2} + 16 \lambda_{{3}} \lambda_{{\Delta 2}} g_{1}^{2} + 256 \lambda_{{2}} \lambda_{{\Delta 1}} g_{2}^{2}\\
				&- 16 \lambda_{1}^{2} \lambda_{{3}} - 42 \lambda_{{3}} g_{1}^{2} g_{2}^{2} + \frac{1235}{16} \lambda_{{2}} g_{2}^{4}\\
				&- 32 \lambda_{1} \lambda_{{3}} \lambda_{{2}} + 72 \lambda_{1} \lambda_{{2}} g_{2}^{2} - 70 \lambda_{{\Delta 2}}^{2} \lambda_{{2}}\\
				&+48 \lambda_{{3}} \lambda_{{\Delta 1}} g_{1}^{2} + \frac{185}{4} \lambda_{{3}} g_{2}^{4} + 20 \lambda_{{\Delta 1}} g_{1}^{2} g_{2}^{2}\\
				&- 16 \lambda_{{3}} \lambda_{{\Delta 2}}^{2} + 56 \lambda_{{3}} \lambda_{{\Delta 2}} g_{2}^{2} - 118 g_{1}^{2} g_{2}^{4}\\
				&- 13 \lambda_{{2}}^{3} + 5 \lambda_{{2}}^{2} g_{1}^{2} + 11 \lambda_{{2}}^{2} g_{2}^{2}\\
				&+36 \lambda_{1} \lambda_{{3}} g_{2}^{2} - \frac{11}{2} \lambda_{{3}}^{3} + \frac{11}{4} \lambda_{{3}}^{2} g_{2}^{2}\\
				&- 18 \lambda_{1} \lambda_{{3}}^{2} - 16 \lambda_{{3}} \lambda_{{\Delta 2}} \lambda_{{\Delta 1}} - 48 \lambda_{{3}} \lambda_{{2}} \lambda_{{\Delta 1}}\\
				&+30 \lambda_{1} g_{1}^{4} + 108 \lambda_{{3}} \lambda_{{\Delta 1}} g_{2}^{2} - \frac{1157}{12} g_{1}^{6}\\
				&- 60 \lambda_{1}^{2} \lambda_{{2}} - 24 \lambda_{{3}}^{2} \lambda_{{\Delta 1}} + 192 \lambda_{{\Delta 2}} \lambda_{{2}} g_{2}^{2}\\
				&+\frac{125}{4} \lambda_{{3}} g_{1}^{4} + 96 \lambda_{{\Delta 2}} \lambda_{{2}} g_{1}^{2} - 80 \lambda_{{2}} \lambda_{{\Delta 1}}^{2}\\
				&- 16 \lambda_{{3}} \lambda_{{\Delta 2}} \lambda_{{2}} + 30 \lambda_{{\Delta 2}} g_{1}^{4} - \frac{1877}{12} g_{1}^{4} g_{2}^{2}\\
				&+24 \lambda_{1} \lambda_{{2}} g_{1}^{2} - 120 \lambda_{{\Delta 2}} \lambda_{{2}} \lambda_{{\Delta 1}} + \frac{59}{2} g_{2}^{6}\\
				&+8 \lambda_{1} \lambda_{{3}} g_{1}^{2} - 24 \lambda_{{3}} \lambda_{{\Delta 1}}^{2} + \frac{207}{8} \lambda_{{2}} g_{1}^{2} g_{2}^{2}\\\end{align*}\begin{align*}
				\phantom{\left.(4\pi)^2\beta_{\lambda}\right|_{I}=}
				&- 72 \lambda_{1} \lambda_{{2}}^{2} - \frac{39}{4} \lambda_{{3}}^{2} \lambda_{{2}} + 80 \lambda_{{\Delta 1}} g_{2}^{4}\\
				&+60 \lambda_{{\Delta 2}} g_{2}^{4} + 128 \lambda_{{2}} \lambda_{{\Delta 1}} g_{1}^{2} + 40 \lambda_{{\Delta 1}} g_{1}^{4}\\
				&- 18 \lambda_{{3}}^{2} \lambda_{{\Delta 2}} + \frac{17}{4} \lambda_{{3}}^{2} g_{1}^{2} + \frac{6421}{48} \lambda_{{2}} g_{1}^{4}\\
			\end{align*}
			\subsubsection{Evolution of $\lambda_{{\Delta 1}}$}
				\begin{align*}
					\left.(4\pi)^2\beta_{\lambda_{{\Delta 1}}}\right|_{I}=
				&+2 \lambda_{{3}} \lambda_{{2}} + 24 \lambda_{{\Delta 2}} \lambda_{{\Delta 1}} - 12 \lambda_{{\Delta 1}} g_{1}^{2} - 12 g_{1}^{2} g_{2}^{2}\\
				&+6 \lambda_{{\Delta 2}}^{2} + 2 \lambda_{{2}}^{2} + 28 \lambda_{{\Delta 1}}^{2} - 24 \lambda_{{\Delta 1}} g_{2}^{2} + 6 g_{1}^{4} + 15 g_{2}^{4}\\\end{align*}\begin{align*}
			\left.(4\pi)^4\beta_{\lambda_{{\Delta 1}}}\right|_{II}=
				&+12 \lambda_{{3}} \lambda_{{2}} g_{2}^{2} + 24 \lambda_{{\Delta 2}}^{2} g_{1}^{2} + 168 \lambda_{{\Delta 2}} g_{2}^{4}\\
				&+3 \lambda_{{3}}^{2} g_{2}^{2} + 20 \lambda_{{2}} g_{2}^{4} + 352 \lambda_{{\Delta 1}}^{2} g_{2}^{2}\\
				&- 3 \lambda_{{3}}^{2} \lambda_{{\Delta 1}} + 5 \lambda_{{3}} g_{1}^{4} + \frac{466}{3} g_{1}^{4} g_{2}^{2}\\
				&- 284 \lambda_{{\Delta 2}}^{2} \lambda_{{\Delta 1}} - 20 \lambda_{{2}}^{2} \lambda_{{\Delta 1}} + \frac{841}{3} \lambda_{{\Delta 1}} g_{1}^{4}\\
				&- 96 \lambda_{{\Delta 2}}^{3} + 120 \lambda_{{\Delta 2}}^{2} g_{2}^{2} - 56 \lambda_{{\Delta 1}} g_{1}^{2} g_{2}^{2} + 266 \lambda_{{\Delta 1}} g_{2}^{4}\\
				&+4 \lambda_{{3}} \lambda_{{2}} g_{1}^{2} + 192 \lambda_{{\Delta 2}} \lambda_{{\Delta 1}} g_{1}^{2} - 8 \lambda_{{2}}^{3}\\
				&- 6 \lambda_{{3}}^{2} \lambda_{{2}} + 10 \lambda_{{3}} g_{1}^{2} g_{2}^{2} + 120 \lambda_{{\Delta 2}} g_{1}^{4}\\
				&- 12 \lambda_{{3}} \lambda_{{2}}^{2} + 10 \lambda_{{3}} g_{2}^{4} + 10 \lambda_{{2}} g_{1}^{4} - \frac{646}{3} g_{1}^{6}\\
				&- 20 \lambda_{{3}} \lambda_{{2}} \lambda_{{\Delta 1}} + 176 \lambda_{{\Delta 1}}^{2} g_{1}^{2} + 86 g_{1}^{2} g_{2}^{4}\\
				&- 528 \lambda_{{\Delta 2}} \lambda_{{\Delta 1}}^{2} + 4 \lambda_{{2}}^{2} g_{1}^{2} - 384 \lambda_{{\Delta 1}}^{3}\\
				&- 4 \lambda_{{3}}^{2} \lambda_{{\Delta 2}} - 144 \lambda_{{\Delta 2}} g_{1}^{2} g_{2}^{2} + 12 \lambda_{{2}}^{2} g_{2}^{2}\\
				&- \lambda_{{3}}^{3} + 384 \lambda_{{\Delta 2}} \lambda_{{\Delta 1}} g_{2}^{2} - 209 g_{2}^{6}\\
			\end{align*}
			\subsubsection{Evolution of $\lambda_{{\Delta 2}}$}
				\begin{align*}
					\left.(4\pi)^2\beta_{\lambda_{{\Delta 2}}}\right|_{I}=
				&+18 \lambda_{{\Delta 2}}^{2} + 24 \lambda_{{\Delta 2}} \lambda_{{\Delta 1}} - 12 \lambda_{{\Delta 2}} g_{1}^{2} + 24 g_{1}^{2} g_{2}^{2} - 6 g_{2}^{4}\\
				&+\lambda_{{3}}^{2} - 24 \lambda_{{\Delta 2}} g_{2}^{2}\\\end{align*}\begin{align*}
			\left.(4\pi)^4\beta_{\lambda_{{\Delta 2}}}\right|_{II}=
				&- 7 \lambda_{{3}}^{2} \lambda_{{\Delta 2}} - 20 \lambda_{{\Delta 2}} \lambda_{{2}}^{2} + 96 \lambda_{{\Delta 2}} \lambda_{{\Delta 1}} g_{1}^{2}\\
				&- 4 \lambda_{{3}}^{3} + 2 \lambda_{{3}}^{2} g_{1}^{2} + \frac{361}{3} \lambda_{{\Delta 2}} g_{1}^{4} + 360 \lambda_{{\Delta 2}} g_{1}^{2} g_{2}^{2}\\
				&- 8 \lambda_{{3}}^{2} \lambda_{{\Delta 1}} - 672 \lambda_{{\Delta 2}}^{2} \lambda_{{\Delta 1}} + 144 \lambda_{{\Delta 2}}^{2} g_{2}^{2} + 170 g_{2}^{6}\\
				&- 20 \lambda_{{3}} \lambda_{{\Delta 2}} \lambda_{{2}} + 192 \lambda_{{\Delta 2}} \lambda_{{\Delta 1}} g_{2}^{2} - 14 \lambda_{{\Delta 2}} g_{2}^{4}\\
				&- 8 \lambda_{{3}}^{2} \lambda_{{2}} - 20 \lambda_{{3}} g_{1}^{2} g_{2}^{2} - 412 g_{1}^{2} g_{2}^{4}\\
				&- 228 \lambda_{{\Delta 2}}^{3} + 192 \lambda_{{\Delta 1}} g_{1}^{2} g_{2}^{2} - \frac{1652}{3} g_{1}^{4} g_{2}^{2}\\
				&+120 \lambda_{{\Delta 2}}^{2} g_{1}^{2} - 448 \lambda_{{\Delta 2}} \lambda_{{\Delta 1}}^{2} - 48 \lambda_{{\Delta 1}} g_{2}^{4}\\
			\end{align*}

\newpage
\addcontentsline{toc}{section}{References}
\bibliography{mybibliography}
\bibliographystyle{utphys}

\end{document}